\documentclass[fleqn,usenatbib]{mnras}

\usepackage{newtxtext,newtxmath}
\usepackage[T1]{fontenc}

\DeclareRobustCommand{\VAN}[3]{#2}
\let\VANthebibliography\thebibliography
\def\thebibliography{\DeclareRobustCommand{\VAN}[3]{##3}\VANthebibliography}
\usepackage{graphicx}	
\usepackage{amsmath}
\title[Impact of BALs on quasar redshift errors]{Analysis of the impact of broad absorption lines on quasar redshift measurements with synthetic observations} 

\author[L.A. Garc\'ia et al.]{Luz \'Angela Garc\'ia,$^{1}$\thanks{lgarciap@ecci.edu.co}
Paul Martini,$^{2, 3}$
Alma X. Gonzalez-Morales,$^{4,5}$
Andreu Font-Ribera,$^{6,7}$
\newauthor{Hiram K. Herrera-Alcantar,$^{5}$ Jessica Nicole Aguilar,$^{8}$ Steve Ahlen,$^{9}$ David Brooks,$^{7}$ }
\newauthor{Axel de la Macorra,$^{10}$ Peter Doel,$^{7}$  Jaime E. Forero-Romero,$^{11}$ Julien Guy,$^{8}$}
\newauthor{Theodore Kisner,$^{8}$ Martin Landriau,$^{8}$ Ramon Miquel,$^{6}$ John Moustakas,$^{12}$ Jundan Nie,$^{13}$ }
\newauthor{Claire Poppett,$^{8,14,15}$ Gregory Tarl\'e,$^{16}$ Zhimin Zhou.$^{13}$}
\\\\
$^{1}$Universidad ECCI, Cra. 19 No. 49-20, Bogot\'a, Colombia, C\'odigo Postal 111311\\
$^{2}$ Center for Cosmology and AstroParticle Physics, The Ohio State University, 191 West Woodruff Avenue, Columbus, OH 43210, USA\\
$^{3}$ Department of Astronomy, The Ohio State University, 4055 McPherson Laboratory, 140 W 18th Avenue, Columbus, OH 43210, USA\\
$^{4}$ Consejo Nacional de Ciencia y Tecnolog\'ia, Av. Insurgentes Sur 1582. Colonia Credito Constructor, Del. Benito Jurez C.P. 03940, M\'exico D.F. M\'exico\\
$^{5}$ Departamento de F\'isica, Divisi\'on de Ciencias e Ingenier\'ias, Campus Leon, Universidad de Guanajuato, L\'eon 37150, M\'exico\\
$^{6}$ Institut de F\'{i}sica d'Altes Energies (IFAE), The Barcelona Institute of Science and Technology, Campus UAB, 08193 Bellaterra Barcelona, Spain\\
$^{7}$ Department of Physics \& Astronomy, University College London, Gower Street, London, WC1E 6BT, UK\\
$^{8}$ Lawrence Berkeley National Laboratory, 1 Cyclotron Road, Berkeley, CA 94720, USA\\
$^{9}$ Physics Dept., Boston University, 590 Commonwealth Avenue, Boston, MA 02215, USA\\
$^{10}$ Instituto de F\'{\i}sica, Universidad Nacional Aut\'{o}noma de M\'{e}xico, Cd. de M\'{e}xico C.P. 04510, M\'{e}xico\\
$^{11}$ Departamento de F\'isica, Universidad de los Andes, Cra. 1 No. 18A-10, Edificio Ip, CP 111711, Bogot\'a, Colombia\\
$^{12}$ Department of Physics and Astronomy, Siena College, 515 Loudon Road, Loudonville, NY 12211, USA\\
$^{13}$ National Astronomical Observatories, Chinese Academy of Sciences, A20 Datun Rd., Chaoyang District, Beijing, 100012, P.R. China\\
$^{14}$ Space Sciences Laboratory, University of California, Berkeley, 7 Gauss Way, Berkeley, CA 94720, USA\\
$^{15}$ University of California, Berkeley, 110 Sproul Hall \#5800 Berkeley, CA 94720, USA\\
$^{16}$ University of Michigan, Ann Arbor, MI 48109, USA
}
\date{Accepted XXX. Received YYY; in original form ZZZ}
\pubyear{2023}

\begin{document}
\label{firstpage}
\pagerange{\pageref{firstpage}--\pageref{lastpage}}
\maketitle

\begin{abstract}
Accurate quasar classifications and redshift measurements are increasingly important to precision cosmology experiments. Broad absorption line (BAL) features are present in 15-20\% of all quasars, and these features can introduce systematic redshift errors, and in extreme cases produce misclassifications. We quantitatively investigate the impact of BAL features on quasar classifications and redshift measurements with synthetic spectra that were designed to match observations by the Dark Energy Spectroscopic Instrument (DESI) survey. Over the course of five years, DESI aims to measure spectra for 40 million galaxies and quasars, including nearly three million quasars. Our synthetic quasar spectra match the signal-to-noise ratio and redshift distributions of the first year of DESI observations, and include the same synthetic quasar spectra both with and without BAL features. We demonstrate that masking the locations of the BAL features decreases the redshift errors by about 1\% and reduces the number of catastrophic redshift errors by about 80\%. We conclude that identifying and masking BAL troughs should be a standard part of the redshift determination step for DESI and other large-scale spectroscopic surveys of quasars. 
\end{abstract}


\section{Introduction}
\label{sec:intro}
The Dark Energy Spectroscopic Instrument (DESI) is an ongoing Stage IV ground-based facility focused on studying dark energy and the evolution of structure through baryon acoustic oscillations (BAO) and redshift-space distortions techniques \citep{levi2013}. DESI will provide us with the most extensive redshift map of galaxies and quasars to date \citep{desi2016a,desi2016b,zou2017,dey2019,ruiz2020,zhou2020,yeche2020,raichoor2020,hahn2022,zhou2023,lan2023,myers2023,guy2023,schlegel2023,raichoor2023a,raichoor2023b,schlafly2023,silber2023,miller2023,desi2023a,desi2023b}. The DESI survey successfully measures about 205 quasars per square degree, including 60 deg$^{-2}$ about $z>2.1$ that will include measurements of the Ly$\alpha$ forest \citep{chaussidon22,moustakas2023}. This corresponds to nearly three million quasars, including over 0.8 million at $z>2.1$ within the 14500 deg$^2$ survey footprint. 

The Sloan Digital Sky Survey (SDSS), and especially \textsc{BOSS} (the Baryon Oscillation Spectroscopic Survey) and \textsc{eBOSS} (extended \textsc{BOSS}) surveyed 6000 deg$^2$ and measured more than 500000 quasars in the redshift range of 0.8--3.5. As part of the analysis, different contaminants to the quasar spectra, such as DLAs (Damped Ly$\alpha$ Systems), BALs (Broad Absorption Lines), and other metal absorption lines were identified, and different strategies to mitigate their impact on clustering measurements were implemented \citep{dumasdesbourboux2020}. The last \textsc{SDSS} data release with new e\textsc{BOSS} data (DR16) cataloged more than 750000 quasars, including nearly 100000 BALs \citep{lyke2020}.\newline 

BALs are high column density features in the spectra of quasars produced by clouds of gas moving at high velocities in the quasar host galaxy. Their distance from the black hole is somewhat disputed, as is the mechanism that launches them \citep{ganguly2007,capellupo2011,rodriguez2012}. Regardless of their origin, quasars with BAL features nearly always exhibit absorption on the blue side of the \ion{C}{IV} emission line's systemic redshift, up to 0.1-0.2 of the speed of light \citep{guo2019}, but they are also often associated with many other features such as \ion{Si}{IV}, \ion{N}{V}, \ion{Al}{III}, and \ion{O}{VI} (among other high-ionization metals) and Ly$\alpha$. \newline

Different works focus on understanding BAL through other associated transition lines, claiming that \ion{C}{IV} only provides a lower limit of the BAL's effects. For instance, \cite{hall2012} explores \ion{Si}{IV} and \ion{N}{V} emission lines in the quasar spectra, in addition to the typical search for \ion{C}{IV} to inform about the properties of BAL-quasars. On the other hand, \cite{capellupo2017} diagnose broad absorption lines through the description of powerful outflows associated with \ion{P}{V}. Finally, \cite{chen2020} examine the correlation of BALs in \textsc{SDSS DR12} quasar spectra with absorption lines environments.\newline 
Importantly, these absorption lines generate several issues when present in the Ly$\alpha$ forest: they add noise to the intrinsic signal of the spectrum, thus, induce an incorrect redshift estimate compared with the quasar systemic redshift up to $dz \sim$ 0.01. Also, they absorb a significant amount of the flux blueshifted from the emission line counterpart, and consequently, less secure lines can be used to classify their spectra, leading to wrong spectra diagnostics. Finally, the presence of these broad absorption lines increases the complexity level when estimating the spectra continuum. \newline
 
From early in the \textsc{SDSS} project, the \textsc{BOSS} team created a pipeline to identify BALs by visual inspection, characterize and archive them. For instance, \textsc{SDSS-III} labeled BAL-quasars and removed them from their catalog for Ly$\alpha$ forest analysis \citep{slosar2011}. A similar approach was decided for \textsc{SDSS DR9} \citep{paris2012} and \textsc{SDSS DR12Q} \citep{paris2017, bautista2017}, due to the large uncertainties in the quasar systemic redshift caused by BAL. Finally, \cite{lyke2020} explores an algorithm to identify and analyze \ion{C}{IV}- and \ion{Si}{IV}-BALs in the \textsc{SDSS DR16} catalog. The latter approach opens the possibility of treating these lines instead of eliminating spectra with BALs. More quasars are observed with progressively larger spectroscopic facilities; thus, more BAL-QSOs are also being detected, and discarding valid data is not a smart strategy. In particular, \cite{guo2019} find 16.8\% of BAL-quasars in \textsc{SDSS DR14}. Thus, we seek optimal ways to treat BAL quasars in DESI. For instance, \cite{ennesser2021} explore masking broad absorption lines in e\textsc{BOSS (DR14)} and find that the procedure returns up to 95\% of the total forest pathlength lost in previous surveys and discuss how this strategy impacts the Lyman-$\alpha$ autocorrelation functions. However, even when one can mask the BALs from the spectra, there is another effect to consider: the error in the redshift estimation due to the BAL presence. This is important for the measurements of the quasar autocorrelation function. In addition,  \citet{youles2022} showed that quasar redshifts uncertainties impact the Lyman $\alpha$ forest auto- and cross-correlation functions.\newline

This work aims to use synthetic spectra to: i) determine the impact of BAL features on quasars redshifts, in particular, those used for Ly$\alpha$ forest studies; ii) quantify the gain on redshift precision by masking the BAL features; iii) determine if masking BALs at the redshift fitting stage is a viable strategy for an experiment like DESI. Throughout this paper, we focus specifically on quasars at $z>$ 1.8. This lower limit is necessary to identify BALs that may be present up to 25000 km$/$s on the blue side of the \ion{C}{IV} line. \newline
The paper is structured as follows: Section~\ref{sec:mocks} describes the simulated spectra and focuses on the introduction and description of BAL features in the mocks. In section~\ref{sec:metho}, we discuss the pipeline and main assumptions. Finally, section~\ref{sec:discussion} builds on this work's results to offer insight on how to treat future DESI data containing BAL features. 

\section{Simulated datasets}
\label{sec:mocks}

The simulated spectra, also referred to as mocks, used in this work were produced in two stages:\newline

\textbf{Raw mocks.} The raw mocks assume a spatially flat $\Lambda$CDM Planck 2015 cosmology \citep{planck2015}, the corresponding mass power spectrum, and a given observed number density of quasars. A quasar catalog is generated by identifying the high-density regions in a Gaussian random field realization and locating quasars in such regions. The set of sightlines from the position of each quasar to an observer's position, what we call skewers, are also drawn from the same gaussian realizations. Subsequently, the skewers are post-processed with \texttt{LyaCoLoRe}, as discussed in \cite{farr2020a}. \texttt{LyaCoLoRe} adds small-scale fluctuations to a gaussian field, then turns this into a physical density used to calculate the optical depth and, finally, the transmitted flux. As a result, for each skewer in the catalog, we have the transmitted flux fraction as a function of wavelength. The position of high-column density lines, such as DLAs, has been identified at this stage. However, for the purpose of this study, we do not take them into account hereafter.\newline

\textbf{Synthetic spectra.} The raw mocks are processed with the codes \texttt{desisim} and \texttt{specsim} \citep{kirkby2016,herrerainprep} to generate a distinct, realistic representation of each quasar spectral energy distribution.  Absorption by BALs, DLAs, other metals in the intergalactic medium (IGM), and the Lyman $\alpha$ forest are then applied to these spectra. However, we prepare mock spectra containing only BALs and no other contaminant. To increase the level of accuracy of the mocks, the code also adds a background QSO continuum, noise and some smoothing of the forest to mimic instrumental resolution to the transmissions. BAL features are added based on a library of 1500 templates constructed by \citet{niu20} based on the BALs identified by \citet{guo2019}.\newline 

Two important properties to characterize BAL quasars are the absorption index AI described by \cite{hall2002} and the balnicity index BI proposed by \cite{weymann1991}. Both of these parameters are computed in the vicinity of the \ion{C}{IV} emission line and are defined as follows:
\begin{equation}\label{ai}
 \text{AI}_{\ion{C}{IV}}=-\int_{25000}^{0}\left[1-\frac{f(v)}{0.9} \right]C(v)dv.
\end{equation}
\noindent AI$_{\ion{C}{IV}}$ is computed from 25000 to 0 km$/$s bluewards the \ion{C}{IV} emission line. The term $f(v)$ is the normalized flux density of the quasar measured with the \ion{C}{IV} line's velocity shift. On the other hand, $C(v)$ is a parameter set to one if the trough extends for more than 450 km/s, and zero otherwise. Finally, the factor 0.9 captures the fact that BAL troughs absorb at least 10\% of the continuum. 

We adopt the associated error to the AI$_{\ion{C}{IV}}$ parameter, $\sigma_{\text{AI}}^2$, as described by \cite{guo2019}:
\begin{equation}\label{sigma_ai}
\sigma_{\text{AI}}^2 = -  \int_{25000}^{0}\left(\frac{\sigma_{f(v)}^2+\sigma_{\text{PCA}}^2}{0.9^2}\right) C(v)dv    
\end{equation}
\noindent where $\sigma_{f(v)}$ corresponds to the flux error in each pixel of the normalized flux density $f(v)$ and $\sigma_{\text{PCA}}$ is the uncertainty found by \cite{guo2019} in their PCA fitting.

The definition of BI$_{\ion{C}{IV}}$ and its error $\sigma_{\text{BI}}^2$ differs from eqs.~\eqref{ai} and \eqref{sigma_ai} by the fact that it only extends to within 3000\,km\,s$^{-1}$ of the line center and the trough has to extend for at least 2000\,km\,s$^{-1}$, rather than for just 450\,km\,s$^{-1}$ as for AI$_{\ion{C}{IV}}$. 
\begin{equation}\label{bi}
 \text{BI}_{\ion{C}{IV}}=-\int_{25000}^{3000}\left[1-\frac{f(v)}{0.9} \right]C(v)dv.  
\end{equation}
The error for BI was introduced by \cite{trump2006}; however, \cite{guo2019} included the additional term $\sigma_{\text{PCA}}$ to account for the error associated with the PCA fitting in their pipeline.
\begin{equation}\label{sigma_bi}
\sigma_{\text{BI}}^2 = - \int_{25000}^{3000}\left(\frac{\sigma_{f(v)}^2+\sigma_{\text{PCA}}^2}{0.9^2}\right) C(v)dv. 
\end{equation}
\noindent Both AI$_{\ion{C}{IV}}$ and BI$_{\ion{C}{IV}}$ parameterize the equivalent width of BAL troughs. The fraction of BAL quasars identified with the AI$_{\ion{C}{IV}}$ $\neq$ 0 criterion is larger than the fraction identified with the BI$_{\ion{C}{IV}}$ criterion because the AI criterion is sensitive to narrower BAL troughs and it extends closer to the line center. 

The balnicity and absorption index distributions of the templates used for the mocks in this work are representative of the full distributions found in \cite{guo2019}.  BAL templates are added multiplicatively to the model quasar spectra before adding the Ly$\alpha$ forest absorption and other features related to the IGM. These simulated spectra are then convolved with a model for the instrument resolution. Lastly, \texttt{specsim} adds noise appropriate to the apparent magnitude of the source and the integration time. \newline

Once \texttt{specsim} is run, three main files are produced for each \texttt{HEALPix} pixel: a truth-, a zbest-, and a spectra-file\footnote{The simulated data is organized by healpy pixels, following the data model for DESI data.}. The first one contains the most relevant information about the quasar and includes the \textit{true} redshift, the number of exposures, and the fluxes and magnitudes used to produce that particular data set. The segment of this file devoted to BALs contains information about the BAL templates that were used to generate the BAL features, including the BAL template ID, the BAL redshift, the AI$_{\ion{C}{IV}}$ and BI$_{\ion{C}{IV}}$ parameters (and their corresponding errors), in addition to the number of distinct components of \ion{C}{IV} with equivalent width larger than 450 km/s, N$_{\ion{C}{IV} 450}$, and the minimum and maximum velocities of the \ion{C}{IV} troughs defined for each one of the N$_{\ion{C}{IV} 450}$ components, $v_{min 450}$ and $v_{max 450}$, respectively.\newline
The zbest file contains a catalog with modified redshifts, including the finger-of-god effect. This intends to emulate the redshift changes that would be introduced when using a redshift fitter. Finally, the spectra file contains the simulated flux for each camera (b,r,z), the inverse variance of the flux, $\sigma^{-2}$, the resolution, and other metadata that are not used in this work. The three files are related through the TARGETID, a unique identifier for each simulated quasar.\newline

In this work, we use two mock realizations:

\begin{itemize}
    \item \textbf{No BAL mock:} Spectra with continuum and Ly$\alpha$ absorption only. These are simulations where the spectra have only the quasar continuum and Ly$\alpha$ absorption but no BALs or other astrophysical effects. We have generated the same realization (same set of spectra) at several multiples of the standard DESI exposure time of 1000s to simulate similar SNR as in the DESI Y1 survey: 1000, 2000, 3000, and 4000 s. Hereafter, we identify this realization with the subscript $_{noBAL}$.\\
    \item \textbf{BAL mock:} Same as the ``No BAL mocks'' except with 16\% of the quasars with BALs. These spectra are identical to the previous case, just with the BAL absorption added. We use these to investigate redshift changes due to the presence of BAL absorption on the same underlying quasar spectra. We have also produced these spectra at several multiples of the standard DESI exposure time (see cases above). We label this realization with the subscript $_{BAL}$.
\end{itemize}

The simulated spectra cover the redshift range from 1.8 to 3.8, and the quasar density of each healpix pixel follows the expected density of quasars for DESI when we began this work (50 quasars per deg$^2$, rather than the 60 deg$^{-2}$ DESI presently achieves). We used a total of 116750 quasar spectra, which constitute the ``No BAL mock''. These same quasars are repeated in the ``BAL mock'', with the exception that 16\% or 18555 have BAL features. We compute the analysis in a subsample of the DESI Y1 expected footprint because running the redshift classifier \texttt{redrock} is computationally expensive. Nonetheless, the subset of spectra is representative of the overall sample.\newline 

\begin{figure*}
\centering 
\includegraphics[scale=0.25]{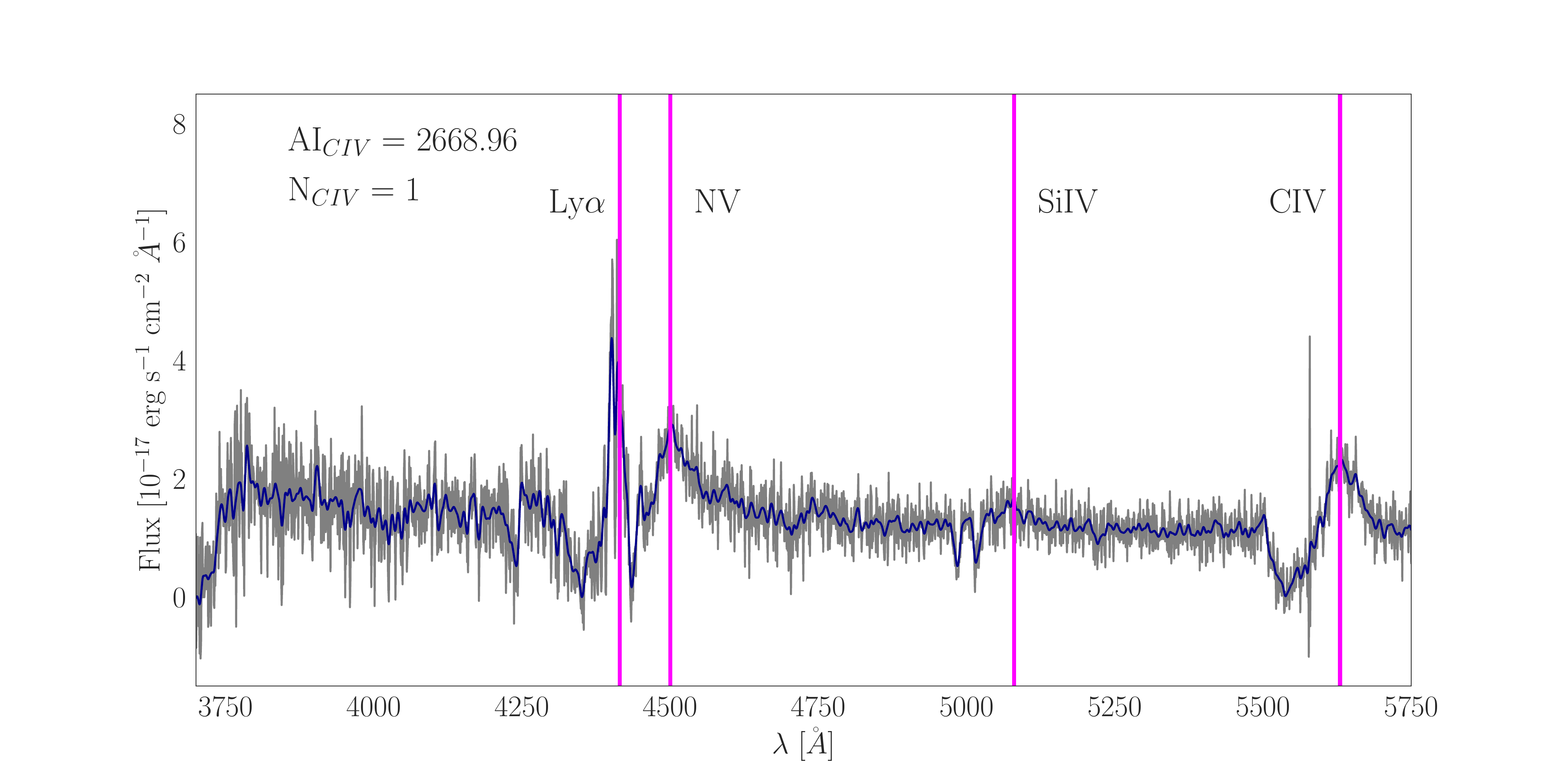}\hspace{0.0cm}
\includegraphics[scale=0.25]{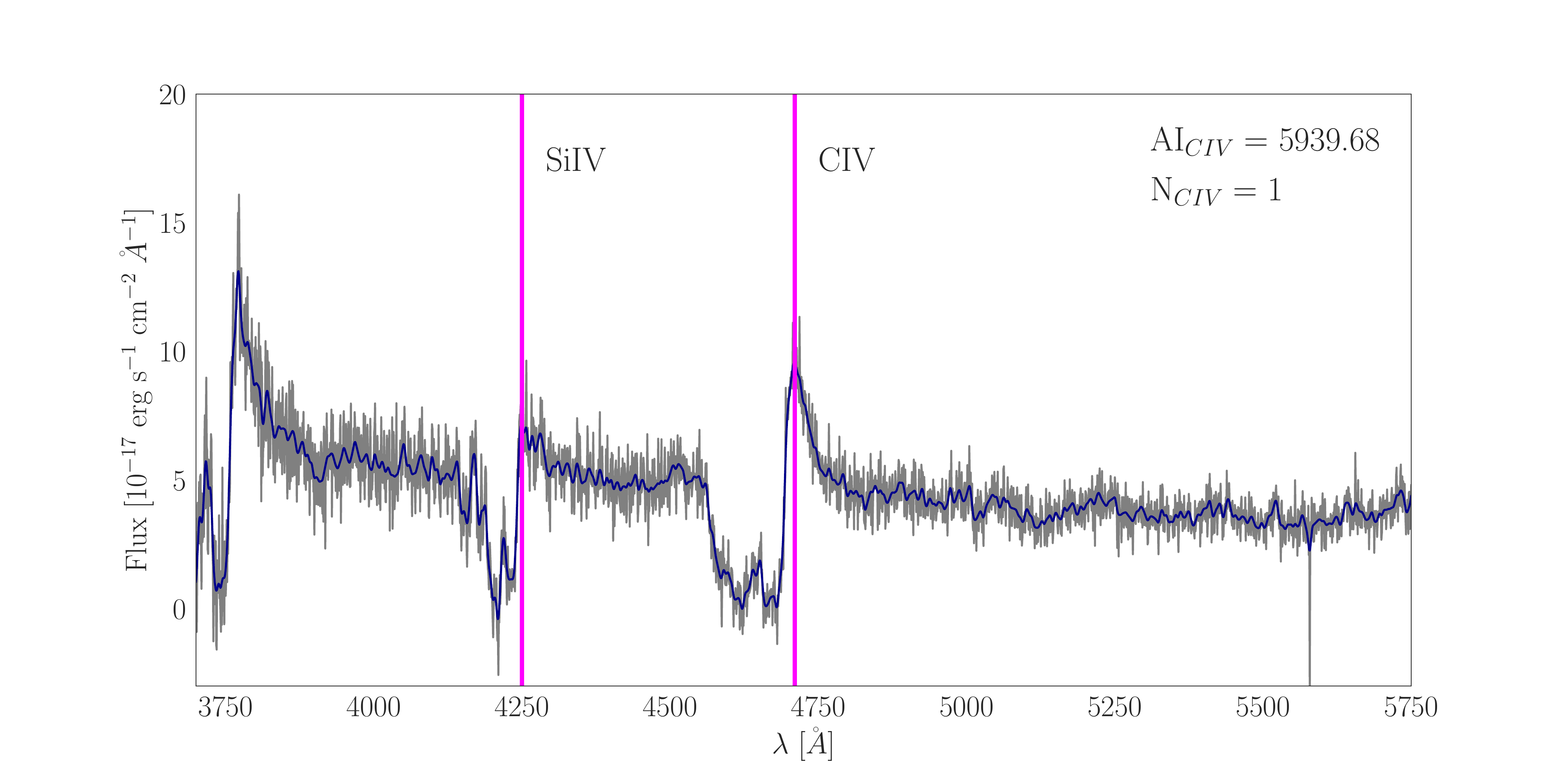}
\caption{\label{fig:i} Examples of synthetic quasar spectra from DESI Y1 mocks as a function of the observed wavelength. The grey region corresponds to the total flux, whereas the dark blue line shows the smoothed flux. The vertical lines show the main lines used to identify the BAL features. In each panel, BAL troughs are exhibited blueward of \ion{C}{IV} emission lines. These spectra' ``true'' synthetic redshifts are 2.632 and 2.039 -upper and lower panels, respectively-. These redshifts are extracted from the truth files of each spectrum.} 
\end{figure*}

Figure~\ref{fig:i} shows two examples of synthetic spectra in our catalog with broad absorption lines blueshifted from the \ion{C}{IV} emission line. Once these troughs are identified, we masked them following the pipeline discussed above.

\section{The effect of masking BALs in quasar spectra}
\label{sec:metho}

We quantify the impact of BALs by measuring redshifts on both mock realizations, one with BALs and one without. Spectral classification and redshift fits for DESI are calculated with the spectral template–redshift fitting code  \texttt{redrock}, which was developed by members of the DESI collaboration\footnote{\url{https://redrock.readthedocs.io/en/latest/api.html}} \citep{redrock2023}. The \texttt{redrock} redshift fitter compares each spectrum against a set of templates for stars, galaxies, and quasars and returns the best match for the input spectrum based on the fit with the minimal $\chi^2$. The return values are a redshift estimate, a class for the type of spectrum fit, and the \texttt{ZWARNING} flag  which is primarily different from 0 for a contaminant (any absorption or skyline that could cause an error in the classification of the spectrum). We run \texttt{redrock} on both realization and study the differences between the output redshifts. We then assess the effect of masking BALs on the redshift fitting to see if there is an improvement with respect to the no-masked BAL case (see details below). \newline 

\begin{figure}
\centering 
\includegraphics[width=1.0\columnwidth]{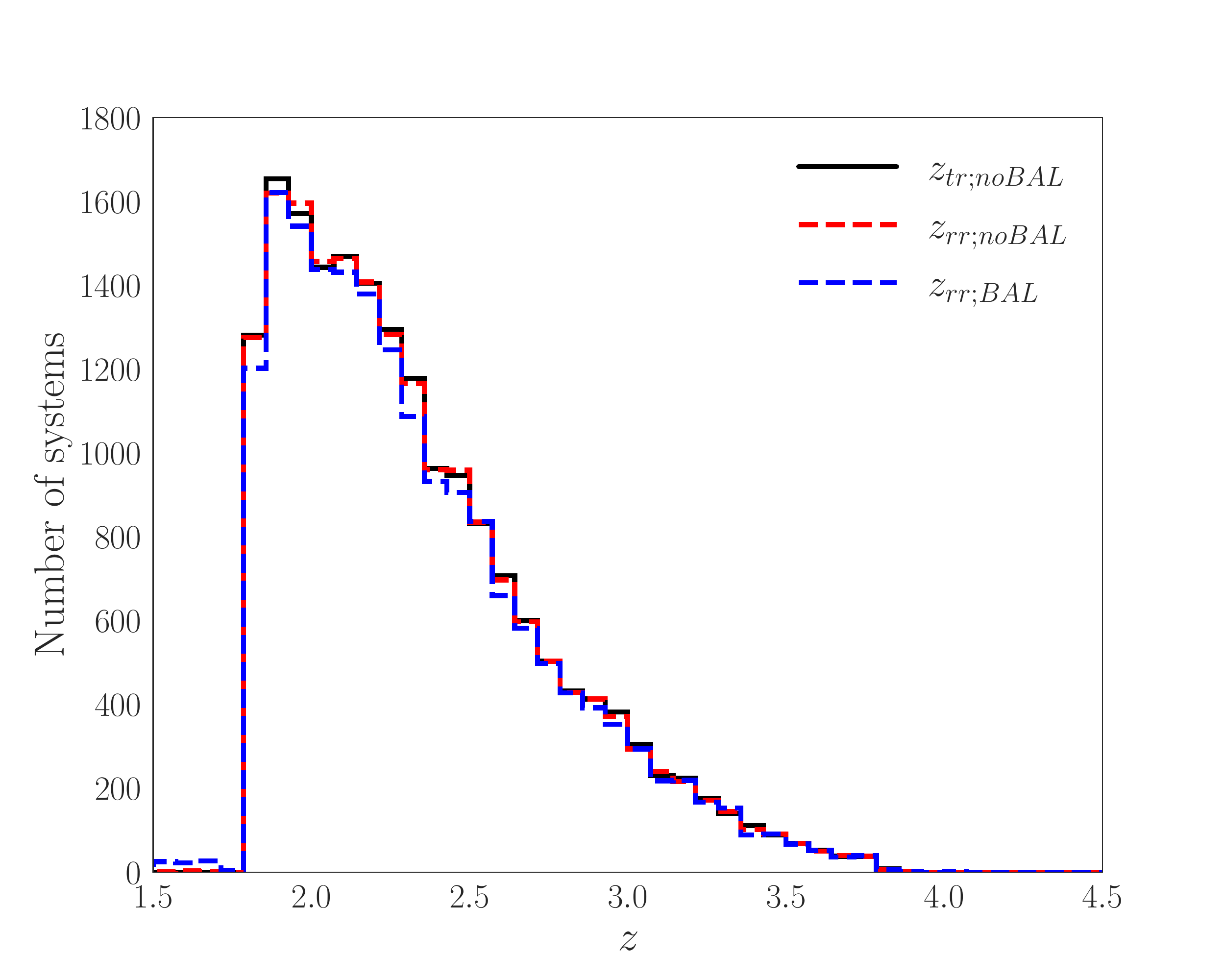}
\caption{\label{fig:ii} Redshift distributions of the 18555 simulated spectra that contain BALs. The histograms show the redshift from the truth file ($z_{tr;noBAL}$, {\it black line}), the measurements from \texttt{redrock} ($z_{rr;noBAL}$; {\it dashed red lines}) for the quasars without the BAL templates; 3) the Ly$\alpha$ $+$ BAL mocks in ({\it blue dashed}). The distributions peak at $z \sim 2$ and decrease at higher redshift.} 
\end{figure}

Figure~\ref{fig:ii} shows three redshift distributions for the quasars: 1) the true redshifts of the quasars $z_{tr;noBAL}$; 2) the \texttt{redrock} redshift distribution for quasars without BALs $z_{rr;noBAL}$; 3) the \texttt{redrock} distribution for quasars with BALs $z_{rr;BAL}$.  The $z_{tr;noBAL}$ redshift is not used in our analysis. However, we draw the comparison here to demonstrate that running \texttt{redrock} introduces a variation in the intrinsic redshift, in addition to the shift caused by the presence of BAL features that induce errors in the redshift estimation. We note that the histograms in Figure~\ref{fig:ii} follow the distribution of the quasars in our catalog, with BAL-quasars being 16 \% of the total number of quasar spectra. While the true redshift range only extends from $1.8 < z < 3.8$, the redshift ranges after running \texttt{redrock} are 0.009 $< z_{rr;noBAL} <$ 5.907, and $-$0.003 $< z_{rr;BAL} <$ 5.907. This is because \texttt{redrock} classifies some of the quasars as stars and in other cases overestimates the redshift by a substantial amount. This problem is more significant for the quasars that are BALs. \newline

\begin{figure*}
\centering 
\includegraphics[scale=0.7]{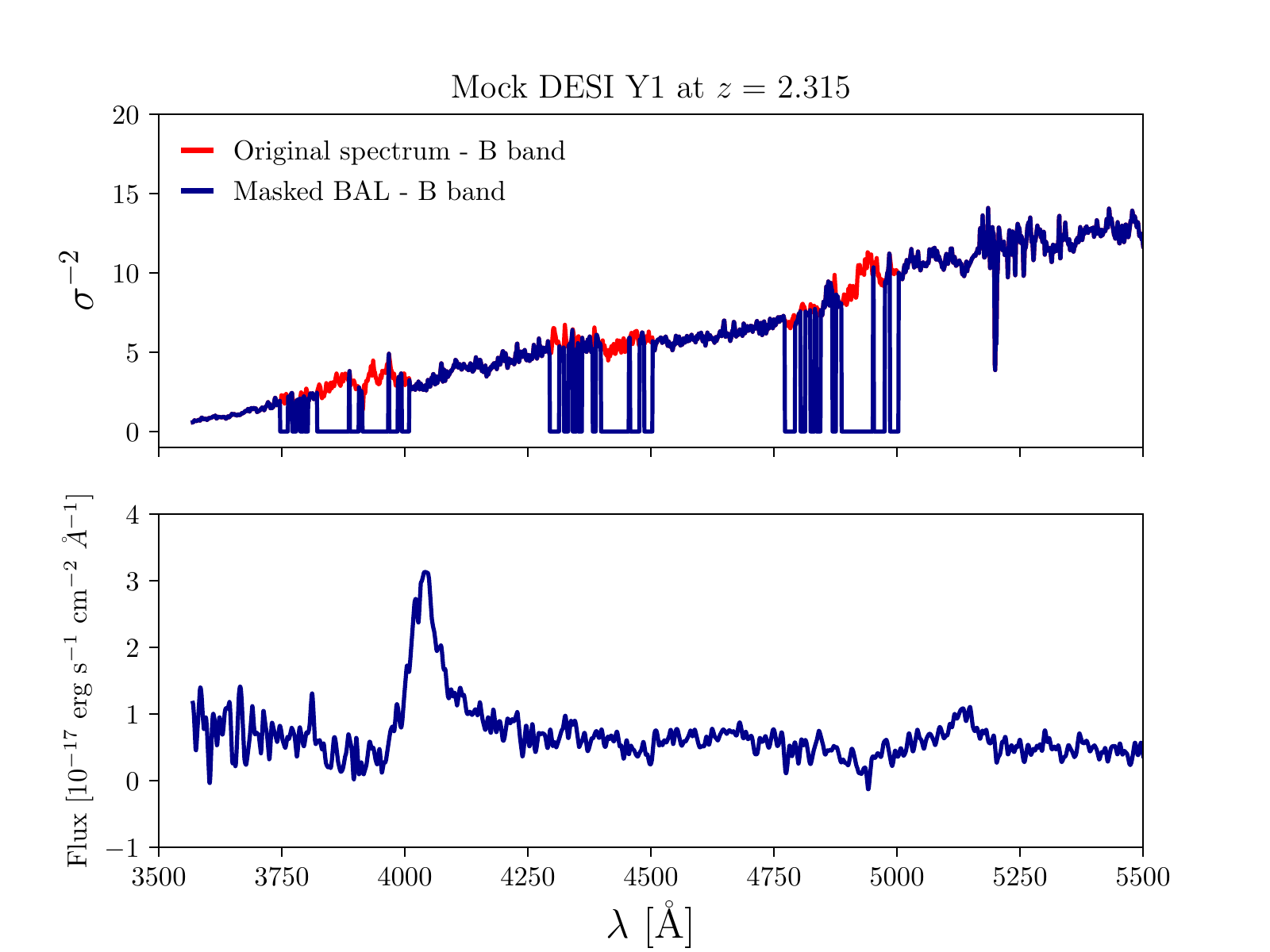}
\caption{\label{fig:via} Inverse variance $\sigma^{-2}$ and smoothed flux of a DESI Y1 synthetic quasar spectra as a function of observed wavelength. The upper panel shows the masked ({\it blue}) and the original inverse variance $\sigma^{-2}$ ({\it red}). When $\sigma^{-2}$ is set to zero, \texttt{redrock} does not fit that part of the spectrum. The lower panel displays the corresponding smoothed flux for each spectrum. The inverse variance set to zero for the eight BAL absorption components on the blue side of \ion{C}{IV}, \ion{Si}{IV}, \ion{N}{V}, and Ly$\alpha$.}
\end{figure*}

We next reran \texttt{redrock} after masking the locations of the BAL features. The information about the locations of the BAL features is stored in a truth catalog that includes the number of troughs associated with the \ion{C}{IV} line that meet the AI criterion N$_{\ion{C}{IV} 450}$ along with the minimum and maximum velocities of each BAL component, $v_{min \ion{C}{IV} 450}$ and $v_{max \ion{C}{IV} 450}$, respectively. Specifically, we determine the observed frame wavelengths that contain each absorption trough based on $v_{min \ion{C}{IV} 450}$ and $v_{max \ion{C}{IV} 450}$ and set the inverse variance of the pixels that contain those wavelengths equal to zero. (The catalog also includes similar information based on the BI criterion, although we do not use that for this study as the trough information based on the AI criterion is more complete.) We also apply the mask to the equivalent wavelength ranges associated with the Ly$\alpha$, \ion{Si}{IV} (1394 \AA) and \ion{N}{V} (1239 \AA) lines. Figure \ref{fig:via} shows the flux and inverse variance of a quasar with and without the BAL features. \newline

\begin{figure*}
\centering
\includegraphics[scale=0.26]{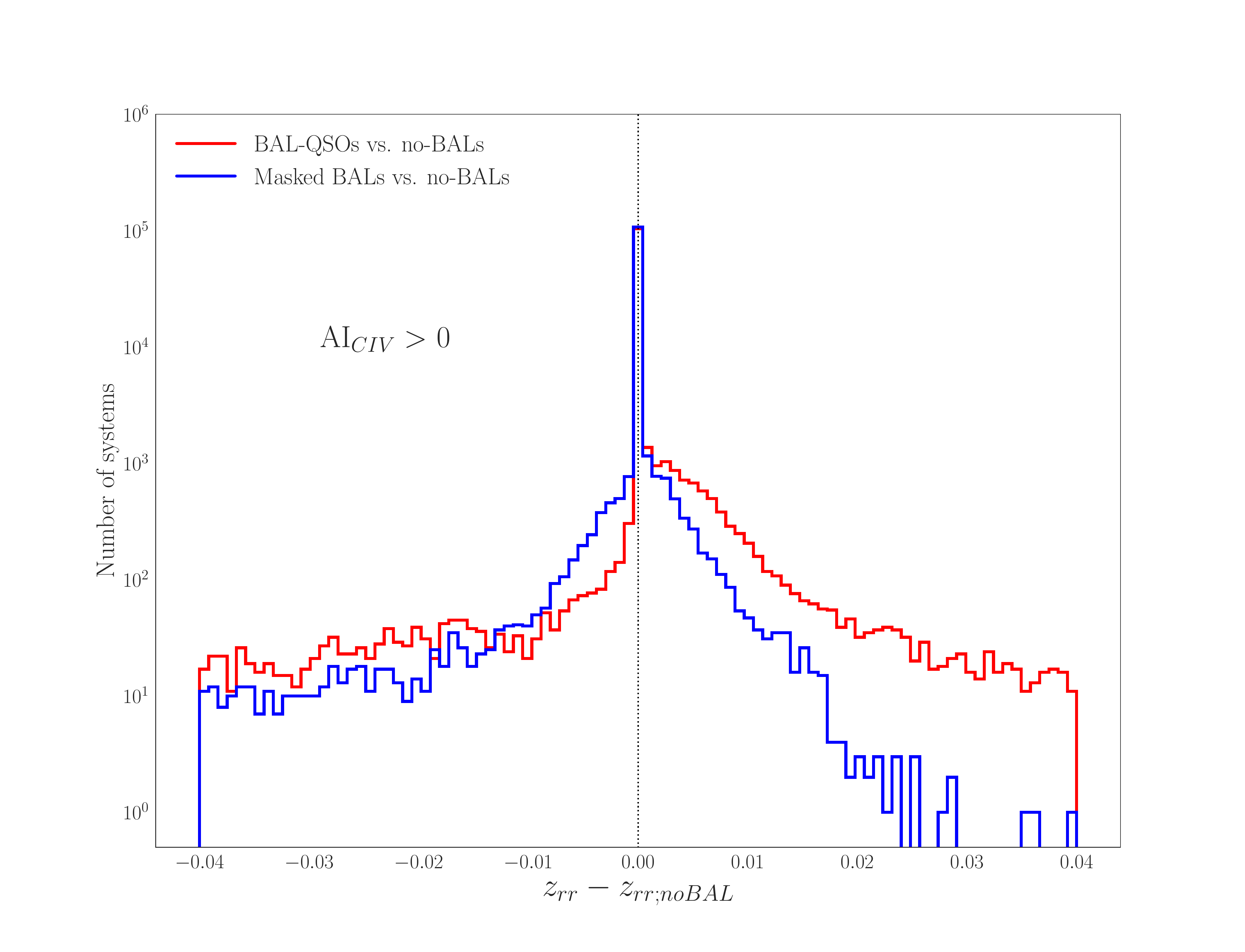}
\caption{\label{fig:1} Distributions of redshift differences for the same quasars with $z_{rr}$ and without BAL features $z_{rr;noBAL}$, both for the case where the BAL features are not masked ({\it red}) and where the BAL features are masked ({\it blue}).  In both cases, the presence of BAL features changes the redshift estimate from \texttt{redrock}, although there are fewer such cases when the BAL features are masked.}
\end{figure*}

Figure~\ref{fig:1} shows the difference between the estimated redshift in the BAL mock, characterized by the absorption index, $\text{AI}_{\ion{C}{IV}} > 0$, and the no BAL mock. We use $z_{rr;noBAL}$ as our \textit{true} redshift, meaning it has not been affected by broad absorption lines or other contaminants, and $z_{rr;BAL}$ is the measured redshift in the sample with BALs (or $z_{rr;mas}$ for masked BAL-QSO). The red line shows the difference for cases where the BAL features are not masked. There are significant redshift changes that indicate the presence of BALs increases the redshift errors. Furthermore, the distribution is asymmetric because the redshifts for the BALs are overestimated relative to the no-BAL sample. This is because the BAL features impact the blue side of the strong emission lines. The blue line in Figure~\ref{fig:1} shows the redshift difference $z_{rr} - z_{rr;noBAL}$ distribution after masking the BAL features. In this case, the distribution is nearly symmetric, and the negative $z_{rr} - z_{rr;noBAL}$ tail is very weak, but still visible in the plot because of the logarithmic scale in the x-axis.\newline

There is an improvement when masks are applied to the BAL features, although a few outliers with large $z_{rr;mas} - z_{rr;noBAL}$ persist. Only 557 out of the 18555 BALs have $|z_{rr;BAL} - z_{rr;noBAL}| > 0.01$ before masking. After masking this number reduces to 103, which corresponds to a reduction of the catastrophic error rate by more than 80\%. These numbers indicate that masking is an excellent approach to reducing the number of catastrophic errors due to the broad absorption lines.\newline 

\begin{figure}
\includegraphics[width=1.1\columnwidth]{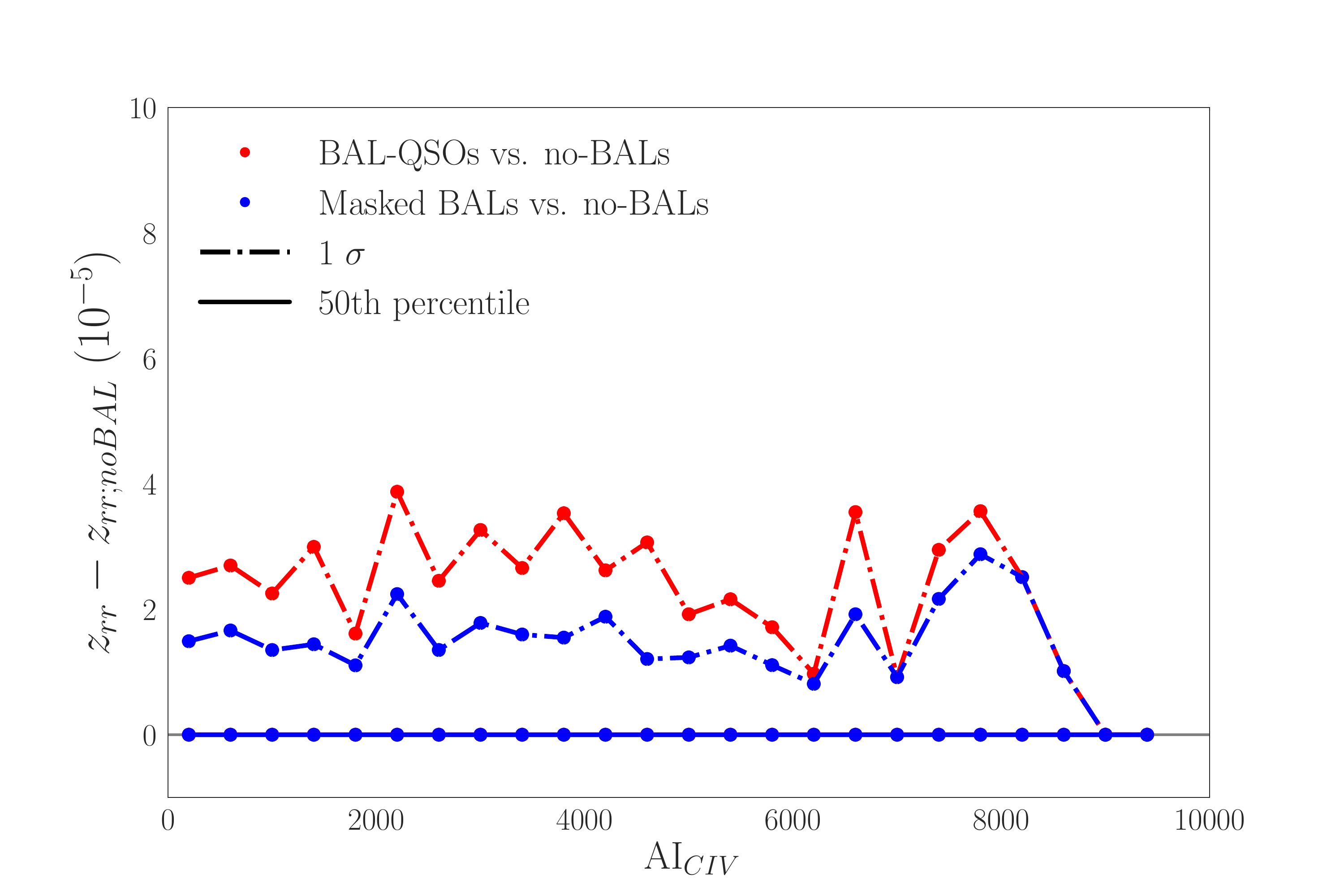}
\caption{\label{fig:ai_diff} Difference between the estimated redshift and the true redshift vs. AI$_{\ion{C}{IV}}$. In this diagram, we present masked BAL with blue points and the original sample of BAL-quasar in red. The 16th (and 84th) and 50th percentiles in the two cases are dashed-dotted and solid lines. Redshift differences in the 16th percentile are overall covered by the 50th percentile.}
\end{figure}

\begin{figure}
\includegraphics[width=1.1\columnwidth]{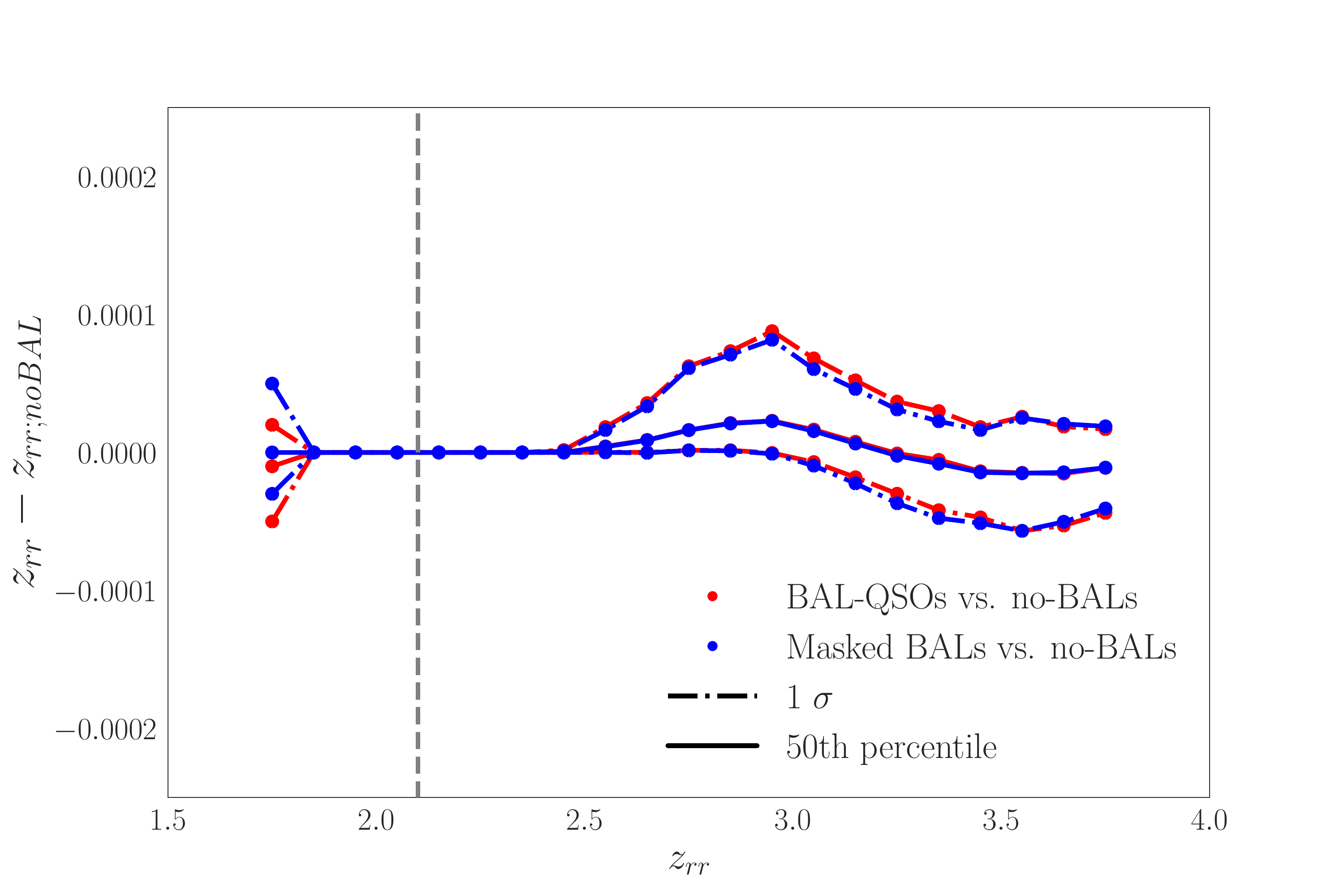}
\caption{\label{fig:z_diff} Difference between the redshift in the absence of BAL features $z_{rr;noBAL}$ compared to when they are present but unmasked ({\it red}) and present and masked ({\it blue}). The samples at $1\sigma$ and the 50th percentile are presented in dashed-dotted and solid lines. The dashed grey vertical line splits the mock sample between tracer and Ly$\alpha$ quasars (left- and right-hand side of the line, respectively). Note that the range of the window is well inside the DESI science requirements.}
\end{figure}

In Figure~\ref{fig:ai_diff}  we display the distribution of $z_{rr} - z_{rr;noBAL}$ as a function of the absorption index AI$_{\ion{C}{IV}}$. The DESI redshift requirements \citep{desi2022} for tracer quasars are 1) the tracer quasar redshift accuracy should be $\sigma_{z} = 0.0025(1+z)$ and 2) a systematic offset on the redshift should be less than $\sigma_{z} = 0.0004(1+z)$. Figures~\ref{fig:ai_diff} and \ref{fig:z_diff} show that $z_{rr} - z_{rr;noBAL}$ is well within the DESI science requirements for both the masked and non-masked BALs.\newline

Figure~\ref{fig:ai_diff} shows that the 50th percentile has a null difference, and the dispersion is roughly constant regardless of the value of AI$_{\ion{C}{IV}}$. This is because even though there is a trend for larger AI values to produce larger redshift errors, most of the BAL features are sufficiently redshifted that they do not have an appreciable impact on the line profiles. 
 
Figure~\ref{fig:z_diff} presents $z_{rr} - z_{rr;noBAL}$ as a function of $z_{rr;noBAL}$. There is no trend with redshift for the vast majority of the sample, with the exceptions at the limits of the redshift range where we identify BALs. At low redshift, there is more scatter due to misclassifications, while above $z >$ 2.5, the \ion{Mg}{II} line is no longer in the spectrograph bandpass, and therefore the \ion{C}{IV} line is more critical for the redshift measurement. \newline

\subsection{Misclassifications and poor fits} \label{sec:class}

The redshift fit performed by \texttt{redrock} also provides the classification corresponding to the minimum $\chi^2$. The output is reflected in Table~\ref{tab:v}. We show the percentage of spectra identified as a quasar, a galaxy, or a star. By construction, \texttt{specsim} simulates only quasar spectra; thus, if it was perfect, \texttt{redrock} should have classified all the spectra in the input sample as quasars. With the no-BAL sample, \texttt{redrock} misidentifies 1.6\% of the spectra and tags them as galaxies. Although \texttt{redrock} does not always diagnose the spectrum as generated by a quasar, there are very few spotted \textit{wrong} cases.
Nonetheless, when 16\% of the synthetic quasars contain BAL features, the number of spectra misidentified as galaxies rises to 3.6\% (the additional 2\%, therefore, originates with the 16\% that are BALs), and one object is classified as a star. In contrast, when the BALs are masked out, the quasar misclassified as a star is no longer present. The number of galaxy-type objects reported by \texttt{redrock} decreases to 2.0\%, close to the misclassification percentage in the absence of BALs. Thus, our masking process cuts down the spectral misclassifications.\newline

\begin{table}
\centering
\caption{Spectral type returned by \texttt{redrock}, comparing synthetic spectra with no BALs, unmasked BALs, and masked BALs. All input spectra are quasars, so the percentage classified as galaxies represents misclassifications. There is an improvement in the number of spectra identified as quasar spectra when the BAL features are masked. A variation in the percentage of the misclassified spectra indicates that the masking procedure considerably improves the performance of \texttt{redrock}. Note the similar results for the no-BAL and masked-BAL samples.}
\label{tab:v} 
\begin{tabular}{|l|c|c|}
\hline
 & \text{Quasar} (\%) & \text{Galaxy} (\%) \\
\hline
\textbf{No BALs}    & 98.4 & 1.6 \\
\textbf{Unmasked BALs} & 96.4 & 3.6 \\
\textbf{Masked BAL} & 98.0 & 2.0 \\ 
\hline
\end{tabular}
\end{table}

We also investigate the rate of redshift warnings \texttt{ZWARNING} and the rate of spectral type misclassifications. For the mock spectra without BALs, we find that 2\% of the spectra have \texttt{ZWARNING} $\neq$ 0, which indicates there is an error associated with the redshift. When BALs are present, this increases to 2.5\%, while after masking the percentage is 2.2\%, very close to the no-BAL value. \newline

We quantify the improvement in the redshift measurements after masking the BAL features with:
\begin{equation}
\label{eq:dv_tr}
dv_{rr} = c \left(\frac{z_{rr}-z_{rr;noBAL}}{1+z_{rr;noBAL}}\right).
\end{equation}
and define the catastrophic error rate as the fraction of systems with $|dv_{rr}| > 15000\,\mathrm{km\,s^{-1}}$ and report the values in Table~\ref{tab:vii}.\newline
\begin{table}
\centering
\caption{Number of systems with absolute values of $|dv_{rr}| > 15000\,\mathrm{km\,s^{-1}}$. We compare a sample of 18555 quasars with unmasked and masked BAL troughs in their spectra. The results reveal that masking BAL troughs effectively reduces cases with $z_{rr} < z_{rr;noBAL}$, despite missing a few dozen cases with estimated redshifts that are greater than the no-BAL case.} 
\label{tab:vii} 
\begin{tabular}{|l|c|c|}
\hline
 & $dv_{rr} < -$ 15000 km$/$s & $dv_{rr} >$ 15000 km$/$s \\
 &  (\%) & (\%) \\
\hline
$z_{rr;BAL} - z_{rr;noBAL}$ & 2.9 & 0.1 \\
$z_{rr;mas} - z_{rr;noBAL}$ & 0.5 & 0.1 \\
\hline
\end{tabular}
\end{table}
The results in Table~\ref{tab:vii} draw three main insights: i) cases with high redshift dispersion occur mostly if BAL features are present. ii) Masking brings down 83\% of the errors due to the presence of BALs in the spectra in the science requirements window. iii) Our masking strategy not only reduces the scatter in the overall quasar sample but also induces a significant decline in the redshift errors for  quasars with the condition $z_{rr} < z_{rr;noBAL}$. The same conclusion does not hold for quasars otherwise. In such case, the numbers stay constant regardless of the masking (reflected in large positive values of $dv_{rr}$ in Table~\ref{tab:vii}). The redshift uncertainty for $dv_{rr} >$ 15000 km$/$s is not boosted by the presence of BAL in the spectrum, but instead, other systematics that are not affected by the masks.\newline 

We define four metrics to investigate further the impact of BALs on the redshift fitting and classification: 
\begin{itemize}
\item \textit{Good fit}: difference in redshift is below a threshold (compared with the \textit{true} redshift $z_{rr;noBAL}$) and \texttt{ZWARNING} $=0$,
\item \textit{Failed fit}: difference in redshift is above a given threshold (compared with the \textit{true} redshift $z_{rr;noBAL}$) and \texttt{ZWARNING} $=0$ (catastrophic failures,
\item \textit{Missed opportunities}: difference in redshift is below a threshold (compared with the \textit{true} redshift $z_{rr;noBAL}$) and \texttt{ZWARNING} $\neq$ 0,
\item \textit{Lost:} difference in redshift is above a threshold (compared with the \textit{true} redshift $z_{rr;noBAL}$) and \texttt{ZWARNING} $\neq$ 0.
\end{itemize}
\noindent The threshold that we use is $\frac{\mid z_{rr}-z_{rr;noBAL}\mid}{z_{rr;noBAL}} = $ 0.05, which is comparable to the allowed tolerance for quasar observations with ground-based telescopes for tracer quasars ($z < 2.1$). The latter is the same threshold assumed to compute results in  Table~\ref{tab:vii}, with the exception that $dv_{rr}$ is expressed in velocity units (a factor of the speed of light, $c$), and this threshold is dimensionless. It presents an error in percentage. Perfect fits in the code would give $dv_{rr} =$ 0 or equivalently, $\frac{\mid z_{rr}-z_{rr;noBAL}\mid}{z_{rr;noBAL}} = $ 0.\newline
In Table~\ref{tab:vi} we report the percentages of quasars in each of these categories for unmasked BALs (top row) and masked BALs (bottom row). The good fits increase by $\sim$ 1.0$\%$ when BAL troughs are covered up; consequently, failed fits and lost opportunities reduce in a similar proportion.\newline

\begin{table}
\centering
\caption{Goodness levels of the fits achieved with \texttt{redrock}. The good fits increase by 1.0\% when BAL features are masked, mostly offset by a corresponding decrease in the percentage of catastrophic errors and lost cases.}
\resizebox{0.48\textwidth}{!}{%
\label{tab:vi} 
\begin{tabular}{|l|c|c|c|c|}
\hline
\text{Fit} & \text{Good} (\%) & \text{Failed} (\%) & \text{Missed} (\%) & \text{Lost} (\%)\\
\hline
$z_{rr;BAL} - z_{rr;noBAL}$ & 97.29 & 0.11 & 2.14 & 0.46 \\ 
$z_{rr;mas} - z_{rr;noBAL}$ & 98.10 & 0.05 & 1.83 & 0.02 \\ \hline
\end{tabular}
}
\end{table}

\begin{figure*}
\centering 
\includegraphics[scale=0.29]{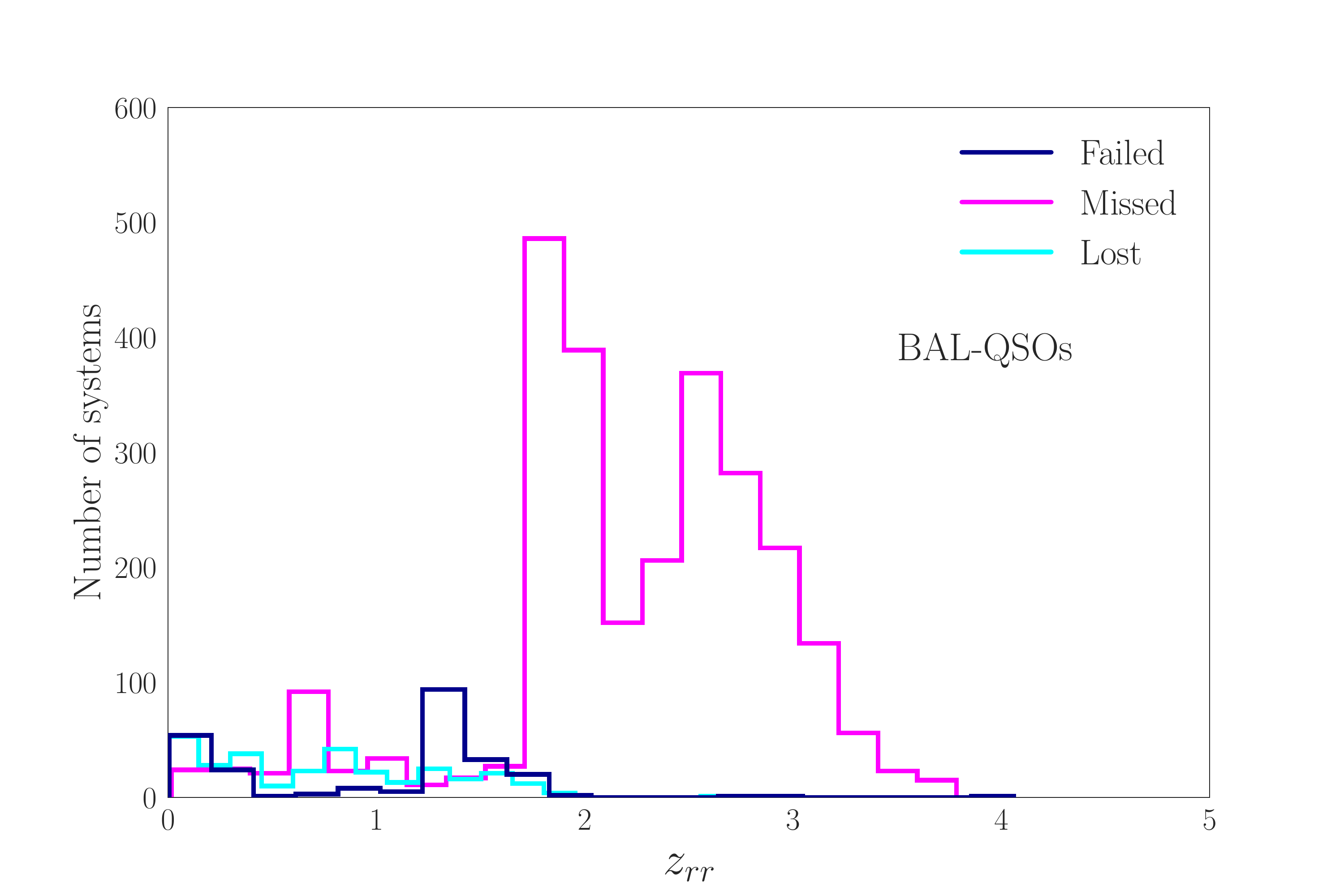}
\includegraphics[scale=0.29]{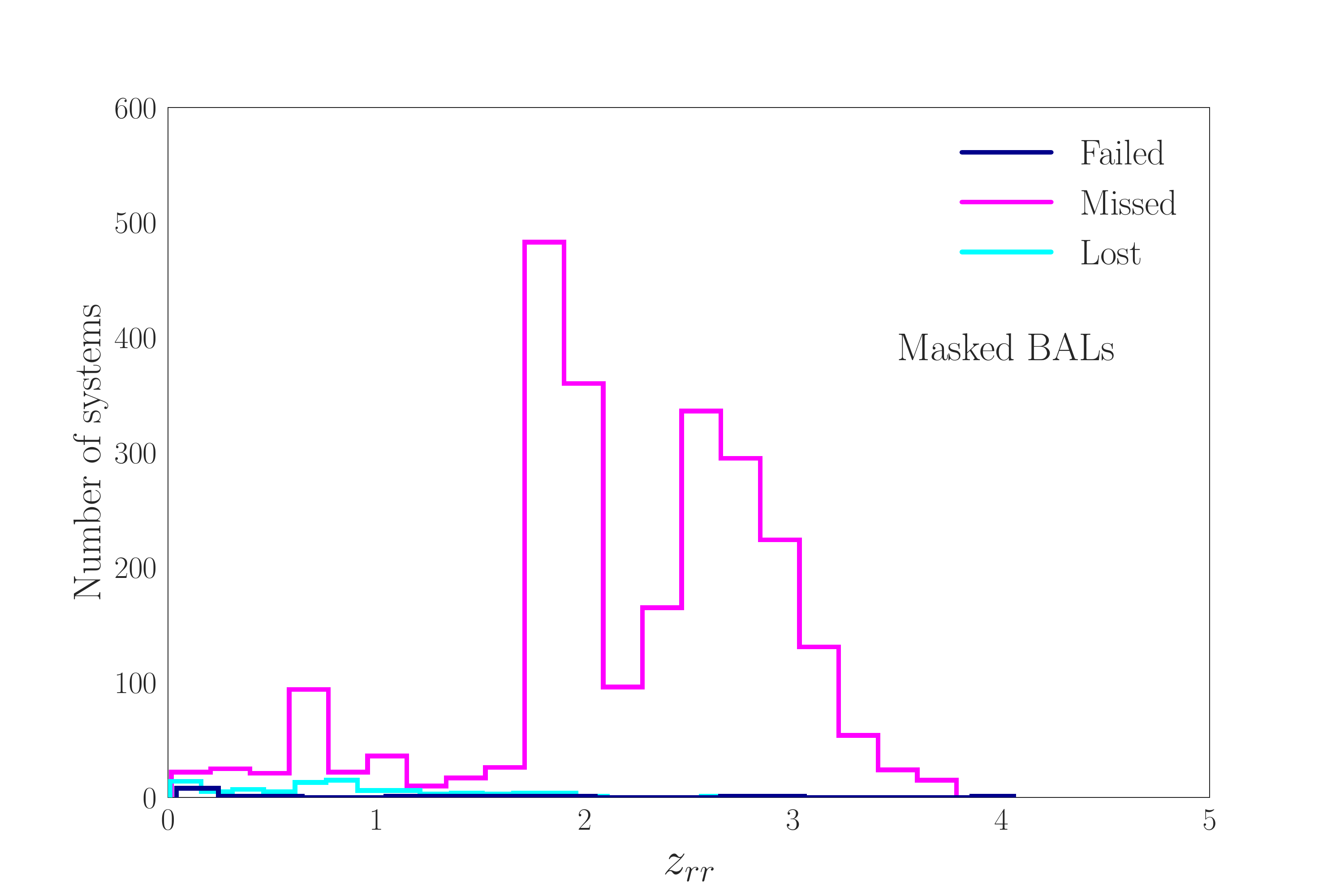}
\caption{\label{fig:v} Distribution of failed fits, missed and lost opportunities with redshift (we skip good fits here since they account for more than 97 \% in all realizations) vs. the redshift of the fit computed by \texttt{redrock}. We compare 116750 quasar realizations with unmasked and masked BAL troughs in the left and right panels.} 
\end{figure*}

Figure~\ref{fig:v} shows the redshift distributions of the Failed, Missed, and Lost categories. The left panel shows the BAL mock without masking compared with the redshift from the mock without BALs, and the right panel shows the same distribution except the BALs are masked. We do not compare these fitting cases with those considered \textit{good} for our pipeline because they outnumbered ``bad'' fits by more than 97-98\%. Figure~\ref{fig:v} shows an interesting trend in that the number of failed and lost fits are mostly present around $z_{rr}\sim$ 1.8. Once the synthetic mocks are extended to lower redshifts, we could investigate this effect in further detail. \newline

When BALs are present in the mocks, the distribution in redshift for the catastrophic failures (dark blue histograms) is primarily seen in the lower-redshift ``tracer'' quasars, with a few occurrences at redshifts above 2.5. On the other hand, missed opportunities (good redshift fitting but \texttt{ZWARNING} $\neq 0$) in magenta lines are centered at $\sim2.0$, and their distribution spans the redshift range of $1.8 - 3.7$. Finally, lost chances (wrong redshift estimate and \texttt{ZWARNING} $\neq 0$) are barely spotted in the masked sample. ``Bad'' fits are largely spotted when BAL occurs in the spectra, and both lost and failed fits scale down when BALs are masked out. The latter results agree with the assumption that led us to run this test: masking the broad absorption lines will reduce the redshift errors in the quasar sample used to study the Ly$\alpha$ forest.

\subsection{Exposure time dependence}\label{sec:expo}

The results presented so far considered mocks designed to represent the data quality for the DESI Year 1 dataset, which corresponds to a nominal exposure time of 1000\,s for all spectra. At the conclusion of the 5-year survey, DESI will observe the $z>2.1$ quasars up to four times (for a nominal exposure time of 4000\,s) to improve the signal-to-noise ratio (SNR) of the Lyman$\alpha$ forest measurement. In this subsection, we investigate the impact of that greater exposure time on the redshift performance. Only quasars with $z >$ 2.1 are candidates for multiple observations, and only a subset of those in our mock data have such longer exposure times. Specifically, the mock data have 41366 and 75384 quasars at $z < 2.1$ and $z \geq 2.1$, respectively. The number of quasars with each exposure time are listed in Table~\ref{tab:xx}. \newline
\begin{table}
\centering
\caption{Effective number of exposures for spectra in the quasar mock sample with BALs. This sample has a total of 116750 quasars, distributed as 41366 tracers quasars ($z <$ 2.1) and 75384 Ly$\alpha$ quasars ($z >$ 2.1).}
\label{tab:xx} 
\begin{tabular}{|l|c|c|c|c|}
\hline
\text{Number of Exposures} & 1000 s & 2000 s & 3000 s & 4000 s\\
\hline
\textbf{Tracer quasars} & 41366 & 0 & 0 & 0\\ 
\textbf{Ly$\alpha$ quasars} & 33371 & 16935 & 11372 & 13706 \\ \hline
\end{tabular}
\end{table}

\begin{figure*}
\centering 
\includegraphics[scale=0.3]{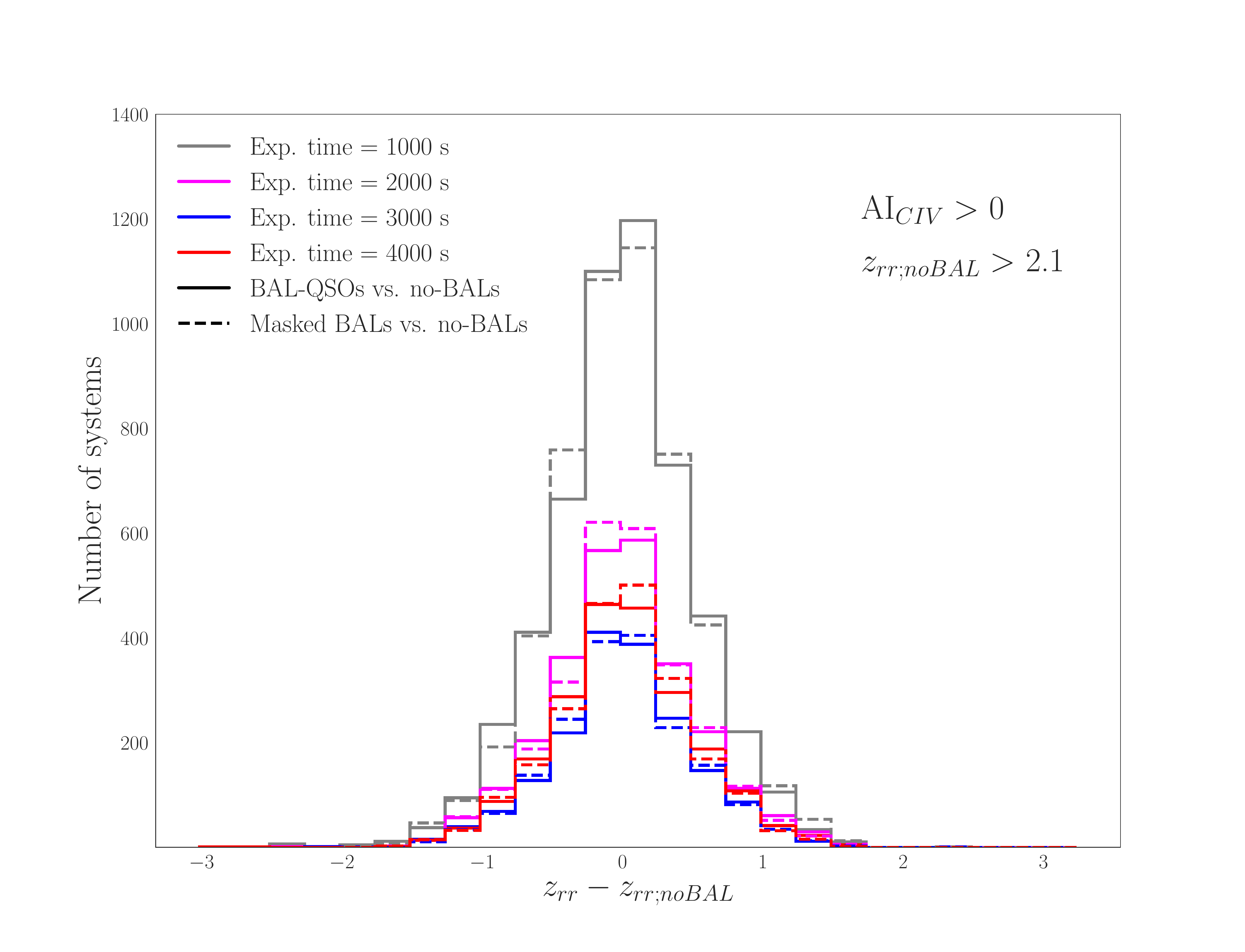}
\caption{\label{fig:all_expt} Distribution $z_{rr} - z_{rr;noBAL}$ for Ly$\alpha$ quasars. The number of exposures displayed is 1000\,s (in grey), 2000\,s (in magenta), 3000\,s (in blue), and 4000\,s (in red). The original BAL and masked BAL samples are presented in solid and dashed lines, respectively.} %{\bf PM: Lastly, it seems like this includes non-BALs too because on the total number. If so, I don't think it is worth showing them as presumably, their redshifts do not change.}}
\end{figure*}

Figure~\ref{fig:all_expt} presents the distribution $z_{rr} - z_{rr;noBAL}$ for quasars with different exposure times. The histograms compare the original sample of BALs in solid lines and masked BALs in dashed lines. The plots exhibit the overall distribution for Ly$\alpha$ quasars ($z_{rr;noBAL} >$ 2.1) with AI$_{CIV} >$ 0 -only BAL quasars, i.e. 18555 in total-; thus, it is clear why many systems have a single exposure, according to Table~\ref{tab:xx}. 

Figure~\ref{fig:all_expt} reinforces the hypothesis of this work: masking out broad absorption lines reduces the discrepancy between the estimated redshift for realizations without and with BALs ($z_{rr;noBAL}$ and $z_{rr;BAL}$). The benefit of masking BALs is also seen for longer exposure times of $2000 - 4000$\,s. However, the increased exposure time does not significantly impact the catastrophic failure rate and the fraction of lost opportunities because those cases are mostly present in the tracer quasar sample. Hence, these differences would not be alleviated with the masking procedure. \newline

In the Appendices, we present additional assessments of the impact of longer exposure times. In Appendix~\ref{app:B}, we perform two additional tests to assess if masking effects in the spectra are related to the specific exposure times. The latter quantity is an indirect measurement of average SNR gain. The results presented in Tables~\ref{tab:x}-\ref{tab:xii} show an improvement in the redshift classification with increasing exposure times when BALs are masked. Yet those results have a very uneven distribution of exposure times, with four times more single-exposure mocks than longer exposure times. We evaluate if this distinction has an impact with a second analysis, presented in Appendix~\ref{app:C}, with the  same number of spectra (11546) for each exposure time: 1000, 2000, 3000, and 4000\,s with no BALs and BALs. The main distinction between the results in the Appendices are that only 16\% of the spectra exhibit BAL features in Appendix~\ref{app:B}, whereas in \ref{app:C}, the percentage of BALs in the spectra is set to a hundred percent.\newline

We draw two important conclusions from these tests: i) the relative distribution of good and ``bad'' fits remains unchanged regardless of the number of exposure times distribution. The most critical cases have a single exposure -an indirect measure of a low SNR in the spectra- that mainly affects tracer quasars. ii) Masking BAL contaminants makes a difference in our synthetic results since this strategy improves the success rate achieved by \texttt{redrock}, regardless of the exposure time of the mock spectra. 

\section{Discussion and conclusions}
\label{sec:discussion}

We have used synthetic quasar spectra to understand the impact of BAL features on quasar redshifts calculated with \texttt{redrock}, the main redshift fitter and object classifier used in DESI.\newline
The first part of the study was devoted to understanding how BAL in the Ly$\alpha$ spectra affect the redshift estimation with the software \texttt{redrock}. Like other absorption lines with large equivalent widths, BAL troughs distort the shape of the spectrum and add noise in the regions with large absorption, which reduces the redshift success rate.\newline

We find that the performance of \texttt{redrock} decreases in several different ways when BALs are present: a small percentage of the synthetic spectra are misclassified (primarily as galaxies, although one as a star), the velocity error $dv$ increases, there are more redshift warnings indicated with the \texttt{ZWARNING} flag, and the percentage of good fits decreases by $\sim$ 0.5\%.\newline

As discussed in \citet{chaussidon22} and \citet{surveyval2022}, there is room for improvement with the fitting procedure performed by \texttt{redrock}. This work demonstrates that masking the BAL regions and re-running the redshift fitter reduces the error in redshift estimation to be nearly comparable to the non-BAL quasars. Specifically, the masking process reduces the number of misidentified objects by more than half and decreases the incidence of redshift warnings \texttt{ZWARNING} $>0$ to only 0.2\% above the 2\% incidence for the non-BAL sample. The redshift efficiency reflected in Table~\ref{tab:vi} shows an improvement at about 1\% for good fits, which results from the combination of fewer misidentified spectra and fewer \texttt{ZWARNING} flags, and a level of scatter that approximates the non-BAL mocks, in particular at low $z$. However, very large redshift dispersions of $dv_{rr} >$ 15000 km$/$s are not corrected even with the masking. \newline

In summary, the broad absorption lines troughs exhibited by $\sim16$\% of quasars introduce redshift errors and contaminate the Lyman$\alpha$ forest region. We have used mock DESI quasar spectra both with and without BAL features to quantify the magnitude of the redshift errors measured with \texttt{redrock} and spectral type misclassifications and catastrophic errors. We have also shown that masking the BAL troughs at the wavelengths of \ion{C}{IV}, \ion{Si}{IV}, \ion{N}{V}, and Lyman$\alpha$ substantially reduces all of these sources of uncertainty and advocate for the automatic identification and masking of the BAL features as part of the quasar identification and redshift fitting process. 

\section*{Acknowledgements}
 L.A. García thanks the Lya WG for allowing her to carry out this project, Universidad ECCI for contributing with funding through the internal allocation v.05--2019, and Jaime Forero-Romero for presenting her into the collaboration. AFR acknowledges support from the Spanish Ministry of Science and Innovation through the program Ramon y Cajal (RYC-2018-025210) and from the European Union’s Horizon
Europe research and innovation programme (COSMO-LYA, grant agreement 101044612). IFAE is partially funded by the CERCA program of the Generalitat de Catalunya. All calculations presented in this work, including the production and storage of the simulated spectra, were done in the supercomputer \texttt{Cori} from the \textsc{National Energy Research Scientific Computing Center (NERSC)} facilities. \newline
This research is supported by the Director, Office of Science, Office of High Energy Physics of the U.S. Department of Energy under Contract No. DE–AC02–05CH11231, and by the National Energy Research Scientific Computing Center, a DOE Office of Science User Facility under the same contract; additional support for DESI is provided by the U.S. National Science Foundation, Division of Astronomical Sciences under Contract No. AST-0950945 to the NSF’s National Optical-Infrared Astronomy Research Laboratory; the Science and Technologies Facilities Council of the United Kingdom; the Gordon and Betty Moore Foundation; the Heising-Simons Foundation; the French Alternative Energies and Atomic Energy Commission (CEA); the National Council of Science and Technology of Mexico (CONACYT); the Ministry of Science and Innovation of Spain (MICINN), and by the DESI Member Institutions: \url{https://www.desi.lbl.gov/collaborating-institutions}.

The authors are honored to be permitted to conduct scientific research on Iolkam Du’ag (Kitt Peak), a mountain with particular significance to the Tohono O’odham Nation.

\bibliographystyle{mnras}
\bibliography{example} 

\begin{thebibliography}{}
\makeatletter
\relax
\def\mn@urlcharsother{\let\do\@makeother \do\$\do\&\do\#\do\^\do\_\do\%\do\~}
\def\mn@doi{\begingroup\mn@urlcharsother \@ifnextchar [ {\mn@doi@}
  {\mn@doi@[]}}
\def\mn@doi@[#1]#2{\def\@tempa{#1}\ifx\@tempa\@empty \href
  {http://dx.doi.org/#2} {doi:#2}\else \href {http://dx.doi.org/#2} {#1}\fi
  \endgroup}
\def\mn@eprint#1#2{\mn@eprint@#1:#2::\@nil}
\def\mn@eprint@arXiv#1{\href {http://arxiv.org/abs/#1} {{\tt arXiv:#1}}}
\def\mn@eprint@dblp#1{\href {http://dblp.uni-trier.de/rec/bibtex/#1.xml}
  {dblp:#1}}
\def\mn@eprint@#1:#2:#3:#4\@nil{\def\@tempa {#1}\def\@tempb {#2}\def\@tempc
  {#3}\ifx \@tempc \@empty \let \@tempc \@tempb \let \@tempb \@tempa \fi \ifx
  \@tempb \@empty \def\@tempb {arXiv}\fi \@ifundefined
  {mn@eprint@\@tempb}{\@tempb:\@tempc}{\expandafter \expandafter \csname
  mn@eprint@\@tempb\endcsname \expandafter{\@tempc}}}

\bibitem[\protect\citeauthoryear{{Abareshi} et~al.,}{{Abareshi}
  et~al.}{2022}]{desi2022}
{Abareshi} B.,  et~al., 2022, \mn@doi [\aj] {10.3847/1538-3881/ac882b}, \href
  {https://ui.adsabs.harvard.edu/abs/2022AJ....164..207A} {164, 207}

\bibitem[\protect\citeauthoryear{{Alexander} et~al.,}{{Alexander}
  et~al.}{2023}]{surveyval2022}
{Alexander} D.~M.,  et~al., 2023, \mn@doi [\aj] {10.3847/1538-3881/acacfc},
  \href {https://ui.adsabs.harvard.edu/abs/2023AJ....165..124A} {165, 124}

\bibitem[\protect\citeauthoryear{{Bailey et al.}}{{Bailey et
  al.}}{2023}]{redrock2023}
{Bailey et al.} 2023

\bibitem[\protect\citeauthoryear{{Bautista} et~al.,}{{Bautista}
  et~al.}{2017}]{bautista2017}
{Bautista} J.~E.,  et~al., 2017, \mn@doi [\aap] {10.1051/0004-6361/201730533},
  \href {https://ui.adsabs.harvard.edu/abs/2017A&A...603A..12B} {603, A12}

\bibitem[\protect\citeauthoryear{{Capellupo}, {Hamann}, {Shields},
  {Rodr{\'\i}guez Hidalgo}  \& {Barlow}}{{Capellupo}
  et~al.}{2011}]{capellupo2011}
{Capellupo} D.~M.,  {Hamann} F.,  {Shields} J.~C.,  {Rodr{\'\i}guez Hidalgo}
  P.,   {Barlow} T.~A.,  2011, \mn@doi [\mnras]
  {10.1111/j.1365-2966.2010.18185.x}, \href
  {https://ui.adsabs.harvard.edu/abs/2011MNRAS.413..908C} {413, 908}

\bibitem[\protect\citeauthoryear{{Capellupo} et~al.,}{{Capellupo}
  et~al.}{2017}]{capellupo2017}
{Capellupo} D.~M.,  et~al., 2017, \mn@doi [\mnras] {10.1093/mnras/stx870},
  \href {https://ui.adsabs.harvard.edu/abs/2017MNRAS.469..323C} {469, 323}

\bibitem[\protect\citeauthoryear{{Chaussidon} et~al.,}{{Chaussidon}
  et~al.}{2023}]{chaussidon22}
{Chaussidon} E.,  et~al., 2023, \mn@doi [\apj] {10.3847/1538-4357/acb3c2},
  \href {https://ui.adsabs.harvard.edu/abs/2023ApJ...944..107C} {944, 107}

\bibitem[\protect\citeauthoryear{{Chen}, {Hamann}, {Ma}, {Lundgren}, {York},
  {Nestor}  \& {AlSayyad}}{{Chen} et~al.}{2020}]{chen2020}
{Chen} C.,  {Hamann} F.,  {Ma} B.,  {Lundgren} B.,  {York} D.,  {Nestor} D.,
  {AlSayyad} Y.,  2020, \mn@doi [\apj] {10.3847/1538-4357/abb401}, \href
  {https://ui.adsabs.harvard.edu/abs/2020ApJ...902...57C} {902, 57}

\bibitem[\protect\citeauthoryear{{DESI Collaboration} et~al.,}{{DESI
  Collaboration} et~al.}{2016a}]{desi2016a}
{DESI Collaboration} et~al., 2016a, arXiv e-prints, \href
  {https://ui.adsabs.harvard.edu/abs/2016arXiv161100036D} {p. arXiv:1611.00036}

\bibitem[\protect\citeauthoryear{{DESI Collaboration} et~al.,}{{DESI
  Collaboration} et~al.}{2016b}]{desi2016b}
{DESI Collaboration} et~al., 2016b, arXiv e-prints, \href
  {https://ui.adsabs.harvard.edu/abs/2016arXiv161100037D} {p. arXiv:1611.00037}

\bibitem[\protect\citeauthoryear{{DESI collaboration}}{{DESI
  collaboration}}{2023a}]{desi2023a}
{DESI collaboration} 2023a

\bibitem[\protect\citeauthoryear{{DESI collaboration}}{{DESI
  collaboration}}{2023b}]{desi2023b}
{DESI collaboration} 2023b

\bibitem[\protect\citeauthoryear{{Dey} et~al.,}{{Dey} et~al.}{2019}]{dey2019}
{Dey} A.,  et~al., 2019, \mn@doi [\aj] {10.3847/1538-3881/ab089d}, \href
  {https://ui.adsabs.harvard.edu/abs/2019AJ....157..168D} {157, 168}

\bibitem[\protect\citeauthoryear{{Ennesser}, {Martini}, {Font-Ribera}  \&
  {P{\'e}rez-R{\`a}fols}}{{Ennesser} et~al.}{2022}]{ennesser2021}
{Ennesser} L.,  {Martini} P.,  {Font-Ribera} A.,   {P{\'e}rez-R{\`a}fols} I.,
  2022, \mn@doi [\mnras] {10.1093/mnras/stac301}, \href
  {https://ui.adsabs.harvard.edu/abs/2022MNRAS.511.3514E} {511, 3514}

\bibitem[\protect\citeauthoryear{{Farr} et~al.,}{{Farr}
  et~al.}{2020}]{farr2020a}
{Farr} J.,  et~al., 2020, \mn@doi [\jcap] {10.1088/1475-7516/2020/03/068},
  \href {https://ui.adsabs.harvard.edu/abs/2020JCAP...03..068F} {2020, 068}

\bibitem[\protect\citeauthoryear{{Ganguly}, {Brotherton}, {Cales}, {Scoggins},
  {Shang}  \& {Vestergaard}}{{Ganguly} et~al.}{2007}]{ganguly2007}
{Ganguly} R.,  {Brotherton} M.~S.,  {Cales} S.,  {Scoggins} B.,  {Shang} Z.,
  {Vestergaard} M.,  2007, \mn@doi [\apj] {10.1086/519759}, \href
  {https://ui.adsabs.harvard.edu/abs/2007ApJ...665..990G} {665, 990}

\bibitem[\protect\citeauthoryear{{Guo} \& {Martini}}{{Guo} \&
  {Martini}}{2019}]{guo2019}
{Guo} Z.,  {Martini} P.,  2019, \mn@doi [\apj] {10.3847/1538-4357/ab2590},
  \href {https://ui.adsabs.harvard.edu/abs/2019ApJ...879...72G} {879, 72}

\bibitem[\protect\citeauthoryear{{Guy} et~al.,}{{Guy} et~al.}{2023}]{guy2023}
{Guy} J.,  et~al., 2023, \mn@doi [\aj] {10.3847/1538-3881/acb212}, \href
  {https://ui.adsabs.harvard.edu/abs/2023AJ....165..144G} {165, 144}

\bibitem[\protect\citeauthoryear{{Hahn} et~al.,}{{Hahn}
  et~al.}{2022}]{hahn2022}
{Hahn} C.,  et~al., 2022, arXiv e-prints, \href
  {https://ui.adsabs.harvard.edu/abs/2022arXiv220808512H} {p. arXiv:2208.08512}

\bibitem[\protect\citeauthoryear{{Hall} et~al.,}{{Hall}
  et~al.}{2002}]{hall2002}
{Hall} P.~B.,  et~al., 2002, \mn@doi [\apjs] {10.1086/340546}, \href
  {https://ui.adsabs.harvard.edu/abs/2002ApJS..141..267H} {141, 267}

\bibitem[\protect\citeauthoryear{{Hall} et~al.,}{{Hall}
  et~al.}{2012}]{hall2012}
{Hall} P.~B.,  et~al., 2012, in {Chartas} G.,  {Hamann} F.,   {Leighly} K.~M.,
  eds,  Astronomical Society of the Pacific Conference Series Vol. 460, AGN
  Winds in Charleston. p.~78

\bibitem[\protect\citeauthoryear{Herrera-Alcantar}{Herrera-Alcantar}{2023}]{herrerainprep}
Herrera-Alcantar H. K. e.~a.,  2023, in preparation

\bibitem[\protect\citeauthoryear{{Kirkby}, {Bailey}, {Guy}  \&
  {Weaver}}{{Kirkby} et~al.}{2016}]{kirkby2016}
{Kirkby} D.,  {Bailey} S.,  {Guy} J.,   {Weaver} B.~A.,  2016, {Quick
  simulations of fiber spectrograph response v0.5}, Zenodo,
  \mn@doi{10.5281/zenodo.154130}

\bibitem[\protect\citeauthoryear{{Lan} et~al.,}{{Lan} et~al.}{2023}]{lan2023}
{Lan} T.-W.,  et~al., 2023, \mn@doi [\apj] {10.3847/1538-4357/aca5fa}, \href
  {https://ui.adsabs.harvard.edu/abs/2023ApJ...943...68L} {943, 68}

\bibitem[\protect\citeauthoryear{{Levi} et~al.,}{{Levi}
  et~al.}{2013}]{levi2013}
{Levi} M.,  et~al., 2013, arXiv e-prints, \href
  {https://ui.adsabs.harvard.edu/abs/2013arXiv1308.0847L} {p. arXiv:1308.0847}

\bibitem[\protect\citeauthoryear{{Lyke} et~al.,}{{Lyke}
  et~al.}{2020}]{lyke2020}
{Lyke} B.~W.,  et~al., 2020, \mn@doi [\apjs] {10.3847/1538-4365/aba623}, \href
  {https://ui.adsabs.harvard.edu/abs/2020ApJS..250....8L} {250, 8}

\bibitem[\protect\citeauthoryear{{Miller et al.}}{{Miller et
  al.}}{2023}]{miller2023}
{Miller et al.} 2023

\bibitem[\protect\citeauthoryear{{Moustakas et al.}}{{Moustakas et
  al.}}{2023}]{moustakas2023}
{Moustakas et al.} 2023

\bibitem[\protect\citeauthoryear{{Myers et al.}}{{Myers et
  al.}}{2023}]{myers2023}
{Myers et al.} 2023

\bibitem[\protect\citeauthoryear{{Niu}}{{Niu}}{2020}]{niu20}
{Niu} W.,  2020, in American Astronomical Society Meeting Abstracts \#235. p.
  108.06

\bibitem[\protect\citeauthoryear{{P{\^a}ris} et~al.,}{{P{\^a}ris}
  et~al.}{2012}]{paris2012}
{P{\^a}ris} I.,  et~al., 2012, \mn@doi [\aap] {10.1051/0004-6361/201220142},
  \href {https://ui.adsabs.harvard.edu/abs/2012A&A...548A..66P} {548, A66}

\bibitem[\protect\citeauthoryear{{P{\^a}ris} et~al.,}{{P{\^a}ris}
  et~al.}{2017}]{paris2017}
{P{\^a}ris} I.,  et~al., 2017, \mn@doi [\aap] {10.1051/0004-6361/201527999},
  \href {https://ui.adsabs.harvard.edu/abs/2017A&A...597A..79P} {597, A79}

\bibitem[\protect\citeauthoryear{{Planck Collaboration} et~al.,}{{Planck
  Collaboration} et~al.}{2016}]{planck2015}
{Planck Collaboration} et~al., 2016, \mn@doi [\aap]
  {10.1051/0004-6361/201525830}, \href
  {https://ui.adsabs.harvard.edu/abs/2016A&A...594A..13P} {594, A13}

\bibitem[\protect\citeauthoryear{{Raichoor} et~al.,}{{Raichoor}
  et~al.}{2020}]{raichoor2020}
{Raichoor} A.,  et~al., 2020, \mn@doi [Research Notes of the American
  Astronomical Society] {10.3847/2515-5172/abc078}, \href
  {https://ui.adsabs.harvard.edu/abs/2020RNAAS...4..180R} {4, 180}

\bibitem[\protect\citeauthoryear{{Raichoor et al.}}{{Raichoor et
  al.}}{2023a}]{raichoor2023a}
{Raichoor et al.} 2023a

\bibitem[\protect\citeauthoryear{{Raichoor} et~al.,}{{Raichoor}
  et~al.}{2023b}]{raichoor2023b}
{Raichoor} A.,  et~al., 2023b, \mn@doi [\aj] {10.3847/1538-3881/acb213}, \href
  {https://ui.adsabs.harvard.edu/abs/2023AJ....165..126R} {165, 126}

\bibitem[\protect\citeauthoryear{{Rodriguez Hidalgo}, {Hamann}, {Eracleous},
  {Capellupo}, {Charlton}  \& {Shields}}{{Rodriguez Hidalgo}
  et~al.}{2012}]{rodriguez2012}
{Rodriguez Hidalgo} P.,  {Hamann} F.,  {Eracleous} M.,  {Capellupo} D.,
  {Charlton} J.,   {Shields} J.,  2012, in {Chartas} G.,  {Hamann} F.,
  {Leighly} K.~M.,  eds,  Astronomical Society of the Pacific Conference Series
  Vol. 460, AGN Winds in Charleston. p.~93 (\mn@eprint {arXiv} {1203.3830})

\bibitem[\protect\citeauthoryear{{Ruiz-Macias} et~al.,}{{Ruiz-Macias}
  et~al.}{2020}]{ruiz2020}
{Ruiz-Macias} O.,  et~al., 2020, \mn@doi [Research Notes of the American
  Astronomical Society] {10.3847/2515-5172/abc25a}, \href
  {https://ui.adsabs.harvard.edu/abs/2020RNAAS...4..187R} {4, 187}

\bibitem[\protect\citeauthoryear{{Schlafly et al.}}{{Schlafly et
  al.}}{2023}]{schlafly2023}
{Schlafly et al.} 2023

\bibitem[\protect\citeauthoryear{{Schlegel et al.}}{{Schlegel et
  al.}}{2023}]{schlegel2023}
{Schlegel et al.} 2023

\bibitem[\protect\citeauthoryear{{Silber} et~al.,}{{Silber}
  et~al.}{2023}]{silber2023}
{Silber} J.~H.,  et~al., 2023, \mn@doi [\aj] {10.3847/1538-3881/ac9ab1}, \href
  {https://ui.adsabs.harvard.edu/abs/2023AJ....165....9S} {p.~9}

\bibitem[\protect\citeauthoryear{{Slosar} et~al.,}{{Slosar}
  et~al.}{2011}]{slosar2011}
{Slosar} A.,  et~al., 2011, \mn@doi [\jcap] {10.1088/1475-7516/2011/09/001},
  \href {https://ui.adsabs.harvard.edu/abs/2011JCAP...09..001S} {2011, 001}

\bibitem[\protect\citeauthoryear{{Trump} et~al.,}{{Trump}
  et~al.}{2006}]{trump2006}
{Trump} J.~R.,  et~al., 2006, \mn@doi [\apjs] {10.1086/503834}, \href
  {https://ui.adsabs.harvard.edu/abs/2006ApJS..165....1T} {165, 1}

\bibitem[\protect\citeauthoryear{{Weymann}, {Morris}, {Foltz}  \&
  {Hewett}}{{Weymann} et~al.}{1991}]{weymann1991}
{Weymann} R.~J.,  {Morris} S.~L.,  {Foltz} C.~B.,   {Hewett} P.~C.,  1991,
  \mn@doi [\apj] {10.1086/170020}, \href
  {https://ui.adsabs.harvard.edu/abs/1991ApJ...373...23W} {373, 23}

\bibitem[\protect\citeauthoryear{{Y{\`e}che} et~al.,}{{Y{\`e}che}
  et~al.}{2020}]{yeche2020}
{Y{\`e}che} C.,  et~al., 2020, \mn@doi [Research Notes of the American
  Astronomical Society] {10.3847/2515-5172/abc01a}, \href
  {https://ui.adsabs.harvard.edu/abs/2020RNAAS...4..179Y} {4, 179}

\bibitem[\protect\citeauthoryear{{Youles} et~al.,}{{Youles}
  et~al.}{2022}]{youles2022}
{Youles} S.,  et~al., 2022, \mn@doi [\mnras] {10.1093/mnras/stac2102}, \href
  {https://ui.adsabs.harvard.edu/abs/2022MNRAS.516..421Y} {516, 421}

\bibitem[\protect\citeauthoryear{{Zhou} et~al.,}{{Zhou}
  et~al.}{2020}]{zhou2020}
{Zhou} R.,  et~al., 2020, \mn@doi [Research Notes of the American Astronomical
  Society] {10.3847/2515-5172/abc0f4}, \href
  {https://ui.adsabs.harvard.edu/abs/2020RNAAS...4..181Z} {4, 181}

\bibitem[\protect\citeauthoryear{{Zhou} et~al.,}{{Zhou}
  et~al.}{2023}]{zhou2023}
{Zhou} R.,  et~al., 2023, \mn@doi [\aj] {10.3847/1538-3881/aca5fb}, \href
  {https://ui.adsabs.harvard.edu/abs/2023AJ....165...58Z} {165, 58}

\bibitem[\protect\citeauthoryear{{Zou} et~al.,}{{Zou} et~al.}{2017}]{zou2017}
{Zou} H.,  et~al., 2017, \mn@doi [\pasp] {10.1088/1538-3873/aa65ba}, \href
  {https://ui.adsabs.harvard.edu/abs/2017PASP..129f4101Z} {129, 064101}

\bibitem[\protect\citeauthoryear{{du Mas des Bourboux} et~al.,}{{du Mas des
  Bourboux} et~al.}{2020}]{dumasdesbourboux2020}
{du Mas des Bourboux} H.,  et~al., 2020, \mn@doi [\apj]
  {10.3847/1538-4357/abb085}, \href
  {https://ui.adsabs.harvard.edu/abs/2020ApJ...901..153D} {901, 153}

\makeatother
\end{thebibliography}

\appendix 
\section{} \label{app:B}

We complement the analysis shown in \S\ref{sec:expo} to assess the performance of \texttt{redrock} by calculating the fits as a function of their time exposures. Table~\ref{tab:x} lists the distribution of spectra identified by \texttt{redrock} in different exposure times. One main conclusion derived from this part of the analysis is that most errors in the spectra classifications occurs for a single exposure. The longer exposure times result is nearly perfect classifications in both the non-BAL and the masked-BAL cases, with only some residual misclassifications in the case of unmasked BALs. 

\begin{table}
\centering
\caption{Spectral type fitted by \texttt{redrock}, comparing synthetic spectra without BALs, unmasked BALs, and masked BALs. We split the entire sample of 116750 quasars in exposure times. Notably, we find that the spectrum fitted as a star (not shown in the Table) has an exposure time of 2000 s and occurs when BALs are added to the synthetic spectra.}
\label{tab:x} 
\begin{tabular}{|lcc|}
\hline
\multicolumn{3}{|c|}{Exposure time $=$ 1000 s}        \\ \hline
\multicolumn{1}{|l|}{}              & \multicolumn{1}{c|}{Quasar (\%)} & Galaxy (\%) \\ \hline
\multicolumn{1}{|l|}{\textbf{No BALs}}  & \multicolumn{1}{c|}{97.5}   & 2.5  \\ 
\multicolumn{1}{|l|}{\textbf{Unmasked BALs}} & \multicolumn{1}{c|}{95.4} &  4.6  \\ 
\multicolumn{1}{|l|}{\textbf{Masked BALs}} & \multicolumn{1}{c|}{96.9} & 3.1 \\ \hline
\end{tabular}
\begin{tabular}{|lcc|}
\hline
\multicolumn{3}{|c|}{Exposure time $=$ 2000 s}  \\ \hline
\multicolumn{1}{|l|}{}  & \multicolumn{1}{c|}{Quasar (\%)} & Galaxy (\%) \\ \hline
\multicolumn{1}{|l|}{\textbf{No BALs}} & \multicolumn{1}{c|}{99.99} & 0.01 \\ 
\multicolumn{1}{|l|}{\textbf{Unmasked BALs}} & \multicolumn{1}{c|}{98.1} & 1.9  \\ 
\multicolumn{1}{|l|}{\textbf{Masked BALs}} & \multicolumn{1}{c|}{99.98} & 0.02 \\ \hline
\end{tabular}
\begin{tabular}{|lcc|}
\hline
\multicolumn{3}{|c|}{Exposure time $=$ 3000 s}  \\ \hline
\multicolumn{1}{|l|}{} & \multicolumn{1}{c|}{Quasar (\%)} & Galaxy (\%) \\ \hline
\multicolumn{1}{|l|}{\textbf{No BALs}} & \multicolumn{1}{c|}{99.99} &  0.01  \\ 
\multicolumn{1}{|l|}{\textbf{Unmasked BALs}} & \multicolumn{1}{c|}{98.0} & 2.0 \\ 
\multicolumn{1}{|l|}{\textbf{Masked BALs}} & \multicolumn{1}{c|}{99.98} & 0.02 \\ \hline
\end{tabular}
\begin{tabular}{|lcc|}
\hline
\multicolumn{3}{|c|}{Exposure time $=$ 4000 s}  \\ \hline
\multicolumn{1}{|l|}{} & \multicolumn{1}{c|}{Quasar (\%)} & Galaxy (\%) \\ \hline
\multicolumn{1}{|l|}{\textbf{No BALs}} & \multicolumn{1}{c|}{100.0} & 0.00\\ 
\multicolumn{1}{|l|}{\textbf{Unmasked BALs}} & \multicolumn{1}{c|}{98.7}  & 1.3 \\ 
\multicolumn{1}{|l|}{\textbf{Masked BALs}} & \multicolumn{1}{c|}{99.99} & 0.01 \\ \hline
\end{tabular}
\end{table}

Table~\ref{tab:xi} shows the percentage of \texttt{ZWARNING} flags when the entire sample of synthetic quasars are split by exposure time. As before, the most significant number of errors occur for a single exposure (1000 s), and the warning flags drop off significantly for more significant exposure times in the observations.

\begin{table}
\centering
\caption{ \texttt{ZWARNING} flags reported by \texttt{redrock} when comparing quasars without BALs, Unmasked BALs and Masked BALs. We split the entire sample of 116750 quasars in different exposure times: 64\% with a single exposure, 14.5\% with 2000 s, 9.7\% with 3000 s, and 11.7\% with 4000 s.}
\label{tab:xi}
\begin{tabular}{|lcc|}
\hline
\multicolumn{3}{|c|}{Exposure time $=$ 1000 s}                                                \\ \hline
\multicolumn{1}{|l|}{}              & \multicolumn{1}{c|}{\texttt{ZWARNING} $=$ 0  (\%)} & \texttt{ZWARNING} $\neq$ 0 (\%) \\ \hline
\multicolumn{1}{|l|}{\textbf{No BALs}} & \multicolumn{1}{c|}{97.25} & 2.75 \\ 
\multicolumn{1}{|l|}{\textbf{Unmasked BALs}} & \multicolumn{1}{c|}{96.68} & 3.32 \\ 
\multicolumn{1}{|l|}{\textbf{Masked BALs}} & \multicolumn{1}{c|}{97.03} & 2.97 \\  \hline
\end{tabular}
\begin{tabular}{|lcc|}
\hline
\multicolumn{3}{|c|}{Exposure time $=$ 2000 s}                                                \\ \hline
\multicolumn{1}{|l|}{} & \multicolumn{1}{c|}{\texttt{ZWARNING} $=$ 0  (\%)} & \texttt{ZWARNING} $\neq$ 0 (\%) \\ \hline
\multicolumn{1}{|l|}{\textbf{No BALs}} & \multicolumn{1}{c|}{99.99}  & 0.01\\ 
\multicolumn{1}{|l|}{\textbf{Unmasked BALs}} & \multicolumn{1}{c|}{99.68}  & 0.32 \\    
\multicolumn{1}{|l|}{\textbf{Masked BALs}} & \multicolumn{1}{c|}{99.97}  &0.03 \\  \hline
\end{tabular}
\begin{tabular}{|lcc|}
\hline
\multicolumn{3}{|c|}{Exposure time $=$ 3000 s}                                                \\ \hline
\multicolumn{1}{|l|}{}   &           \multicolumn{1}{c|}{\texttt{ZWARNING} $=$ 0  (\%)} & \texttt{ZWARNING} $\neq$ 0 (\%) \\ \hline
\multicolumn{1}{|l|}{\textbf{No BALs}} & \multicolumn{1}{c|}{99.54}  & 0.46 \\ 
\multicolumn{1}{|l|}{\textbf{Unmasked BALs}}  & \multicolumn{1}{c|}{99.20}  & 0.80\\ 
\multicolumn{1}{|l|}{\textbf{Masked BALs}} & \multicolumn{1}{c|}{99.38}  & 0.62 \\ \hline
\end{tabular}
\begin{tabular}{|lcc|}
\hline
\multicolumn{3}{|c|}{Exposure time $=$ 4000 s}                                                \\ \hline
\multicolumn{1}{|l|}{}   &           \multicolumn{1}{c|}{\texttt{ZWARNING} $=$ 0  (\%)} & \texttt{ZWARNING} $\neq$ 0 (\%) \\ \hline
\multicolumn{1}{|l|}{\textbf{No BALs}}  & \multicolumn{1}{c|}{100.00}  & 0.0 \\ 
\multicolumn{1}{|l|}{\textbf{Unmasked BALs}}  & \multicolumn{1}{c|}{99.36}  & 0.64\\ 
\multicolumn{1}{|l|}{\textbf{Masked BALs}} & \multicolumn{1}{c|}{99.53}  & 0.47\\ \hline
\end{tabular}
\end{table}

\begin{figure*}
\centering 
\includegraphics[scale=0.275]{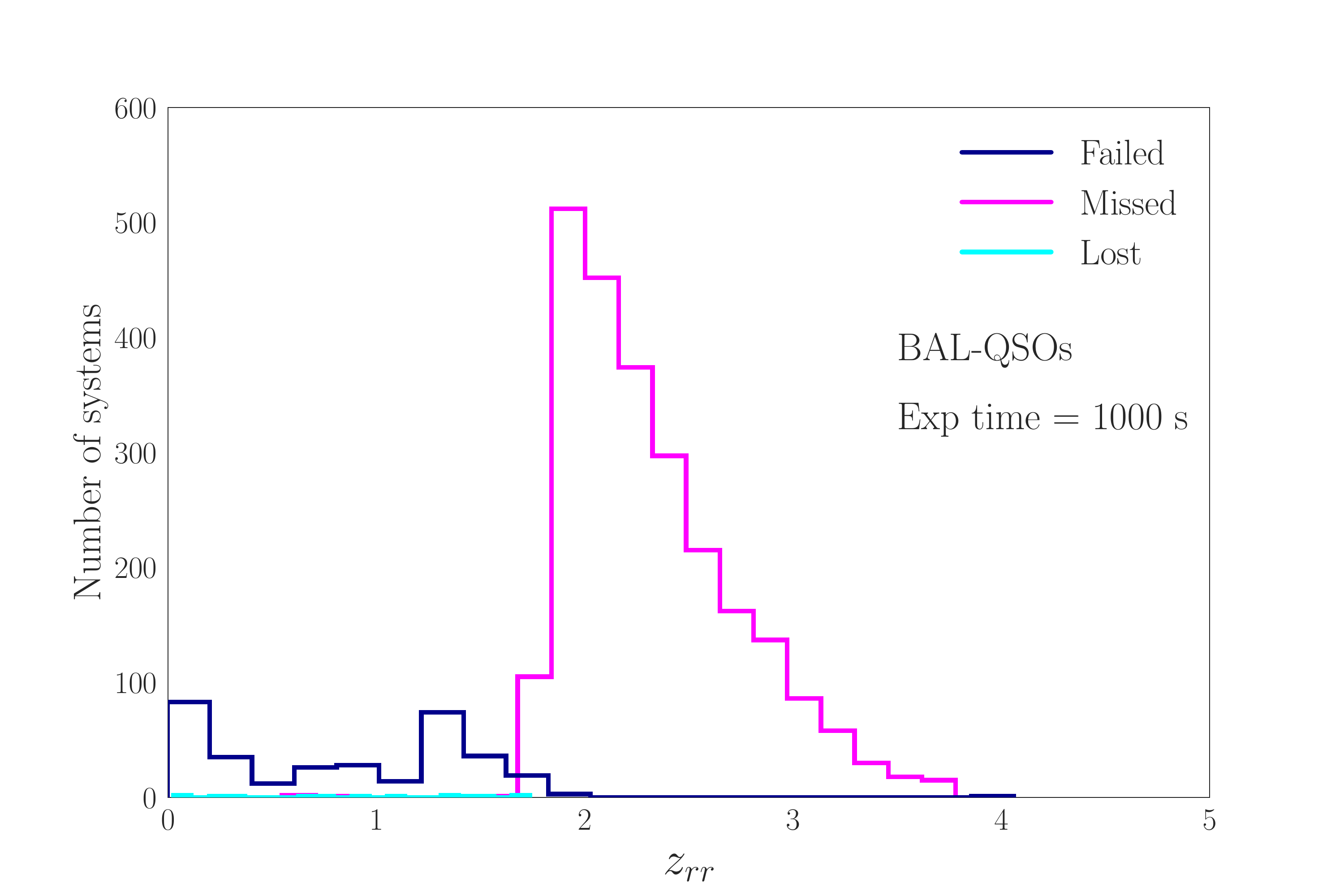}
\includegraphics[scale=0.275]{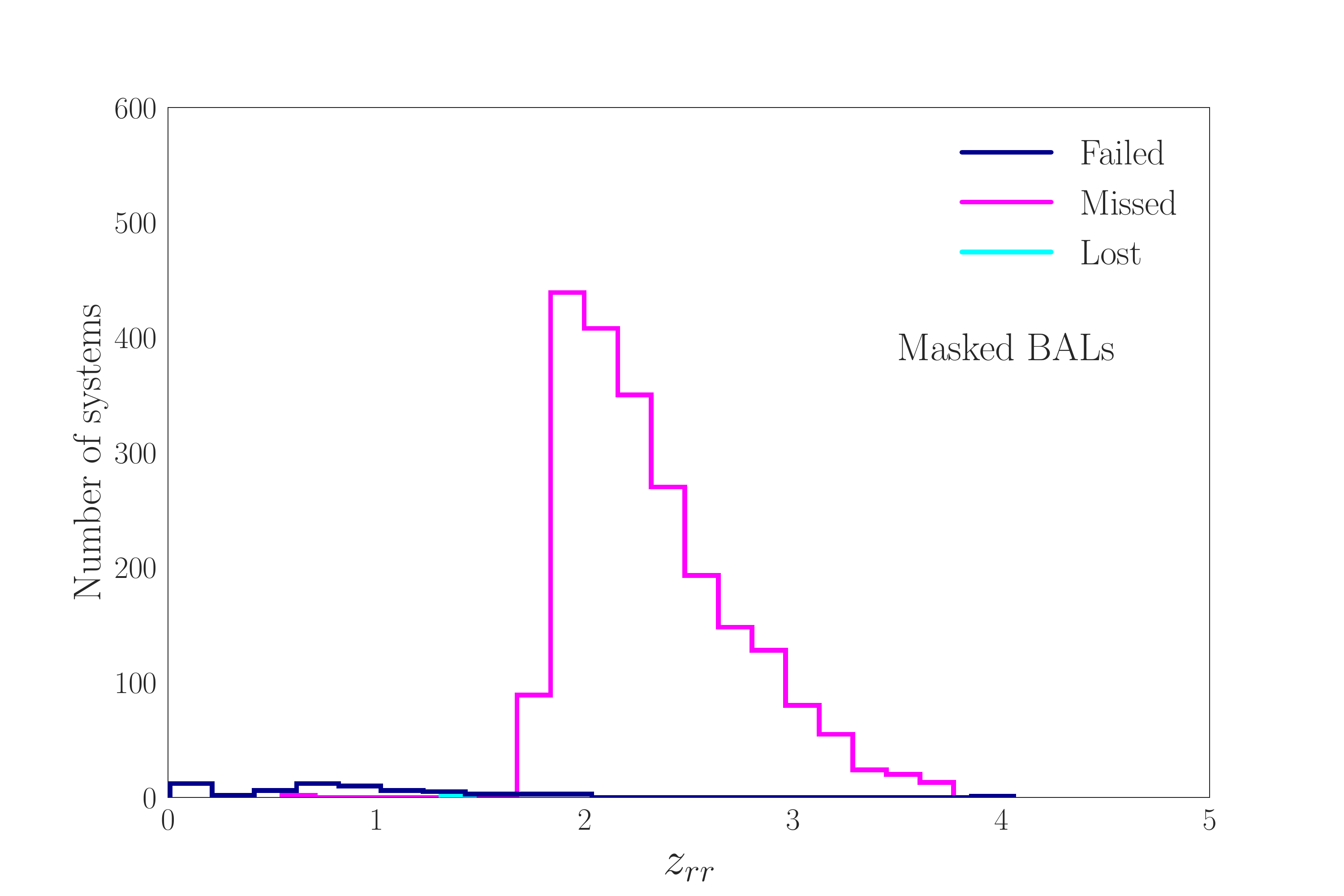}
\vspace{0.0cm}
\hspace{0.0cm}
\includegraphics[scale=0.275]{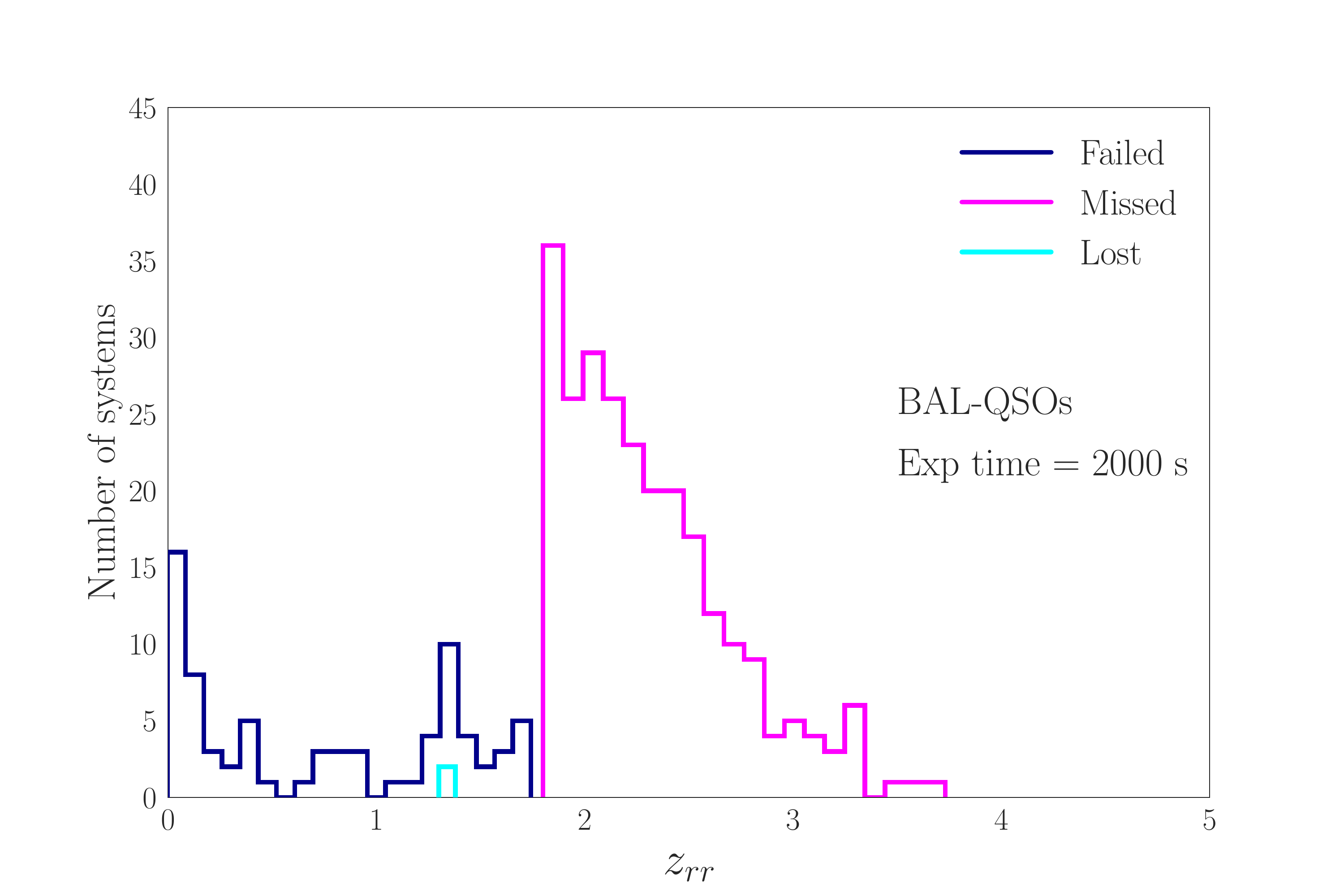}
\includegraphics[scale=0.275]{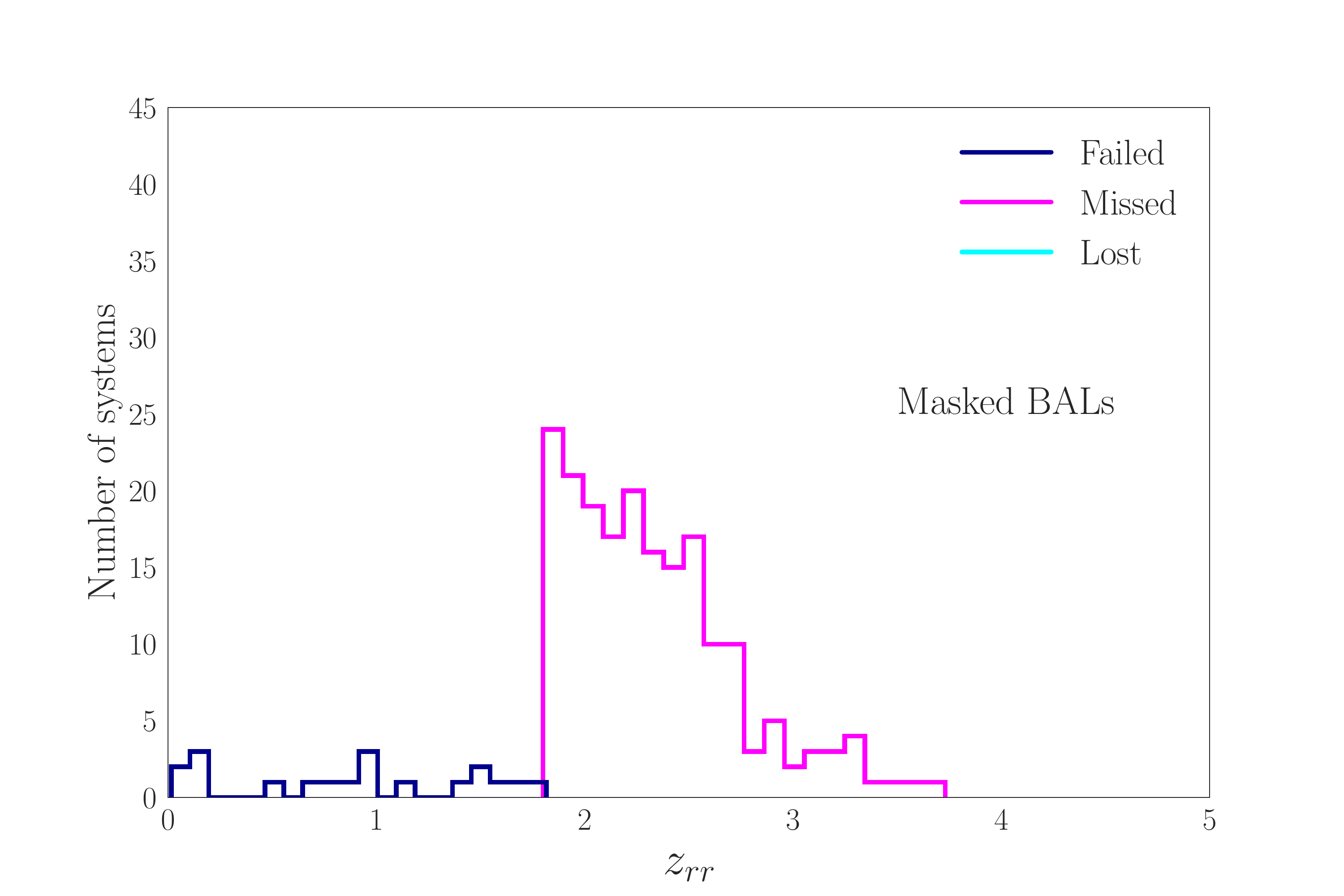}
\vspace{0.0cm}
\hspace{0.0cm}
\includegraphics[scale=0.28]{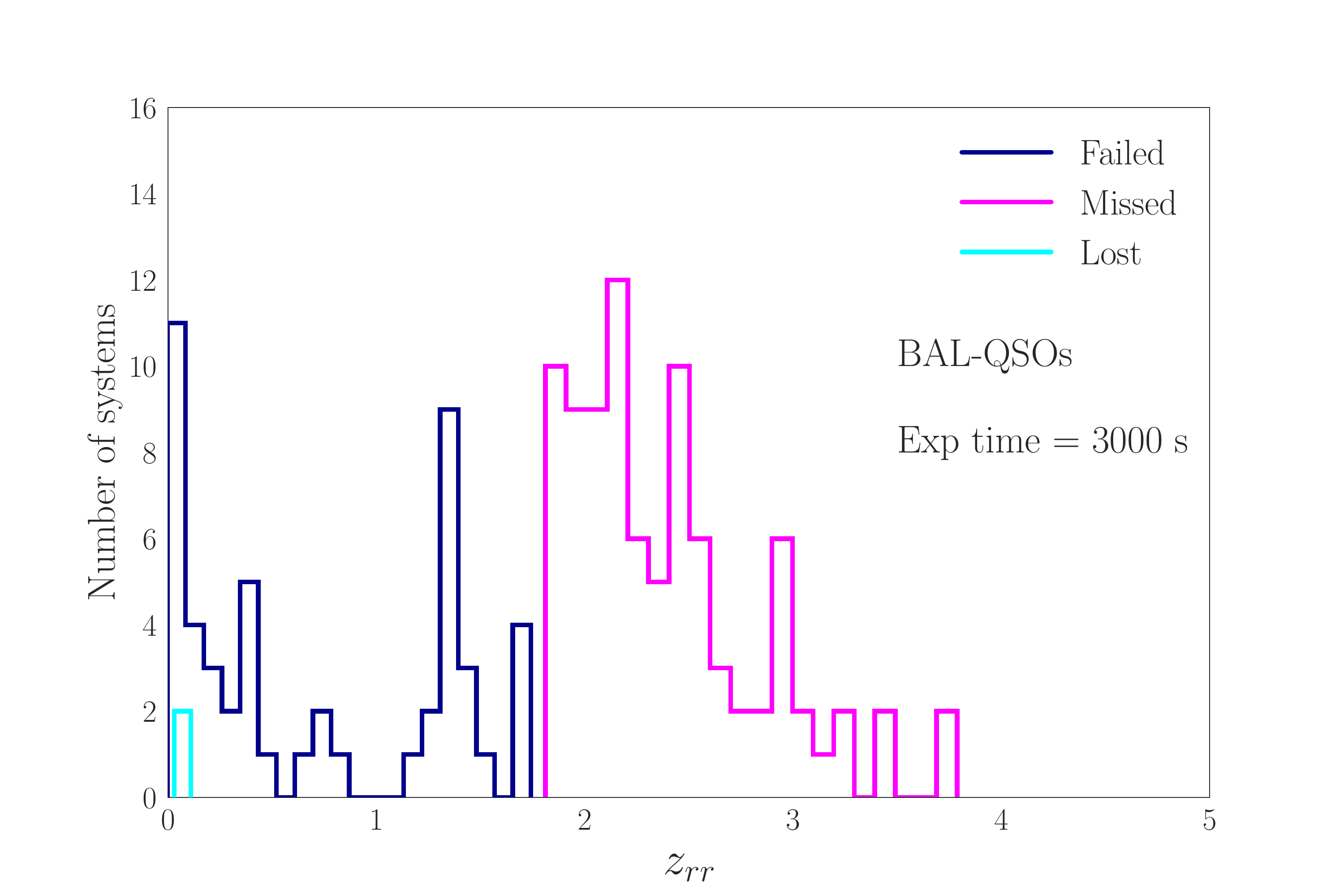}
\includegraphics[scale=0.28]{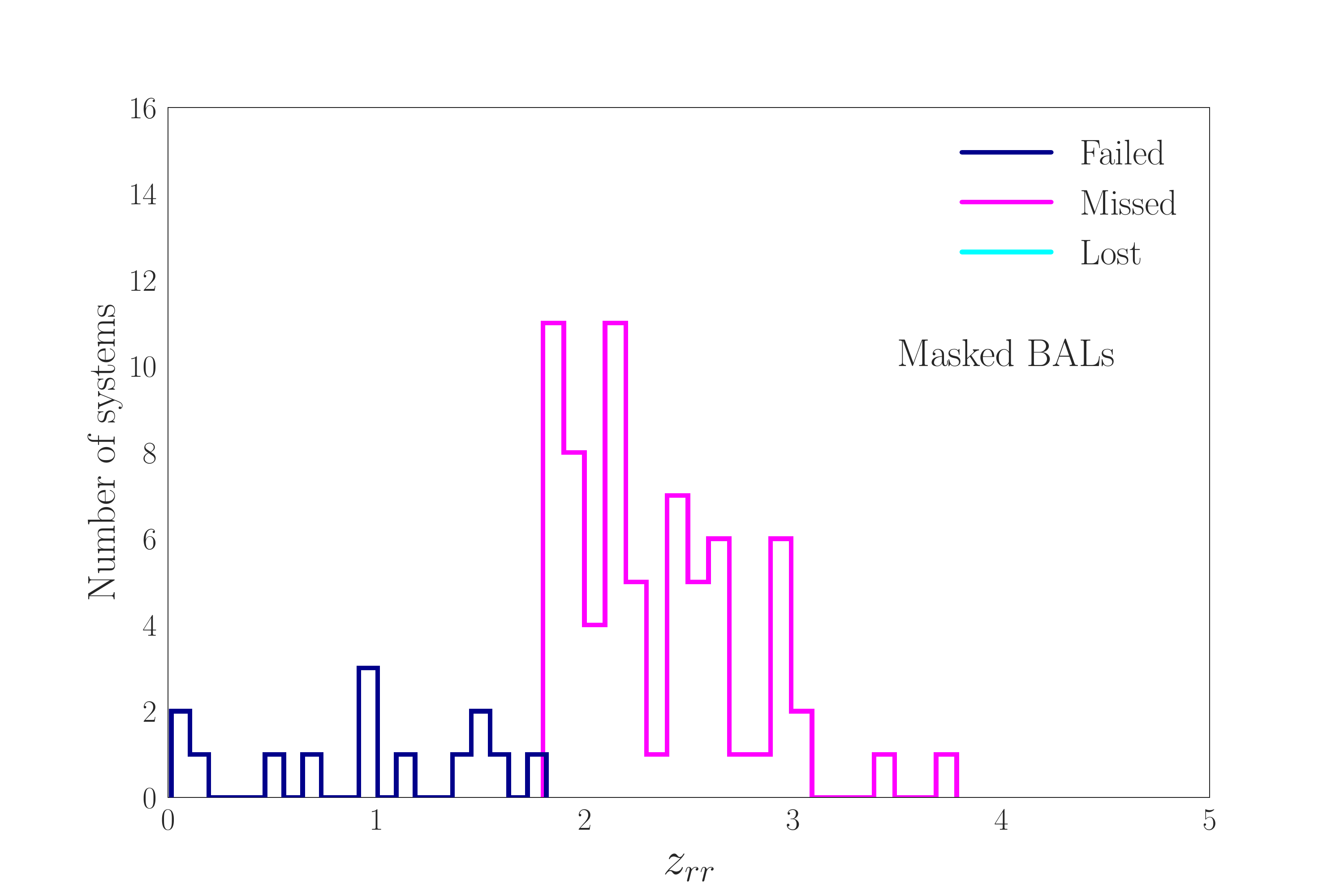}
\vspace{0.0cm}
\hspace{0.0cm}
\includegraphics[scale=0.28]{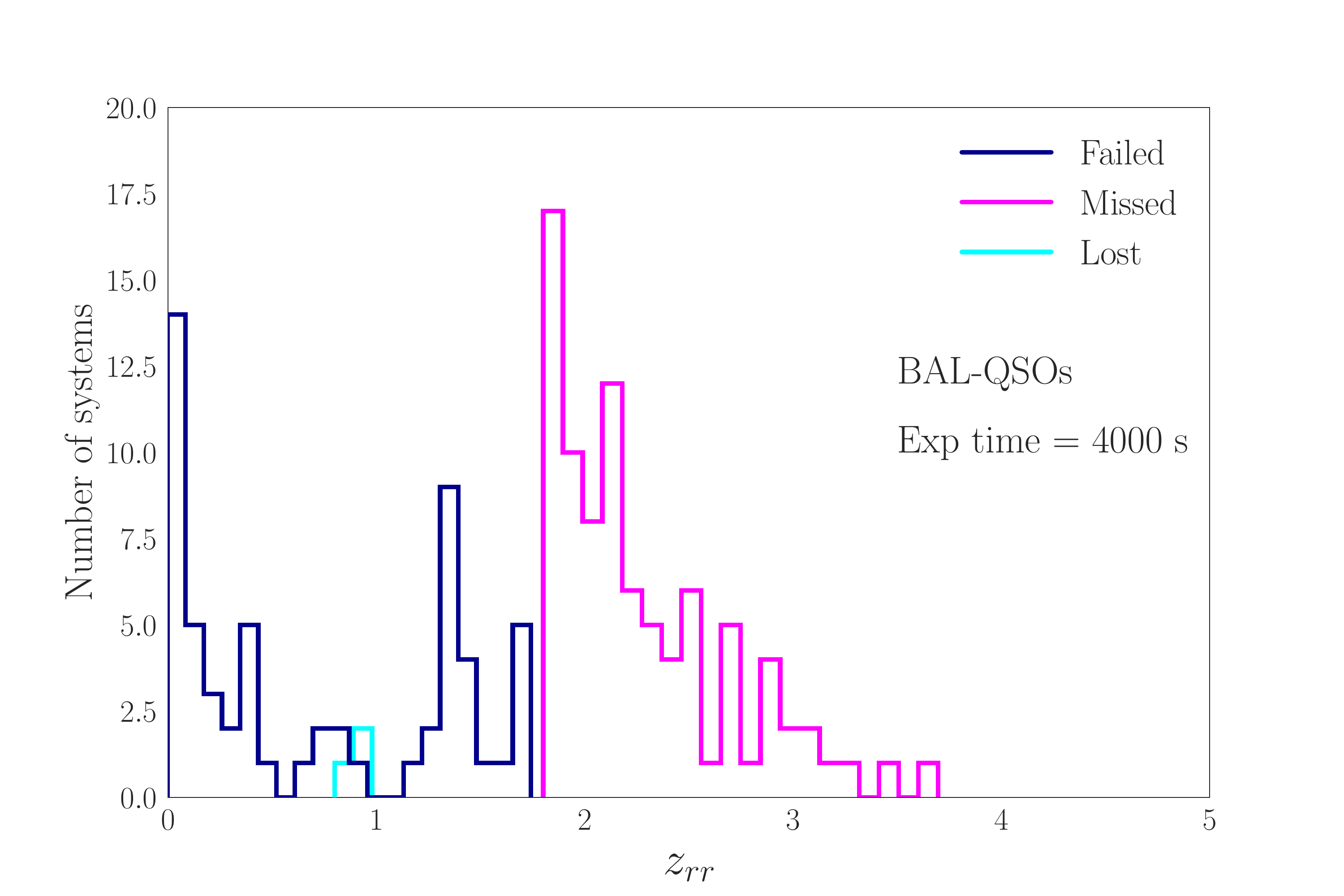}
\includegraphics[scale=0.28]{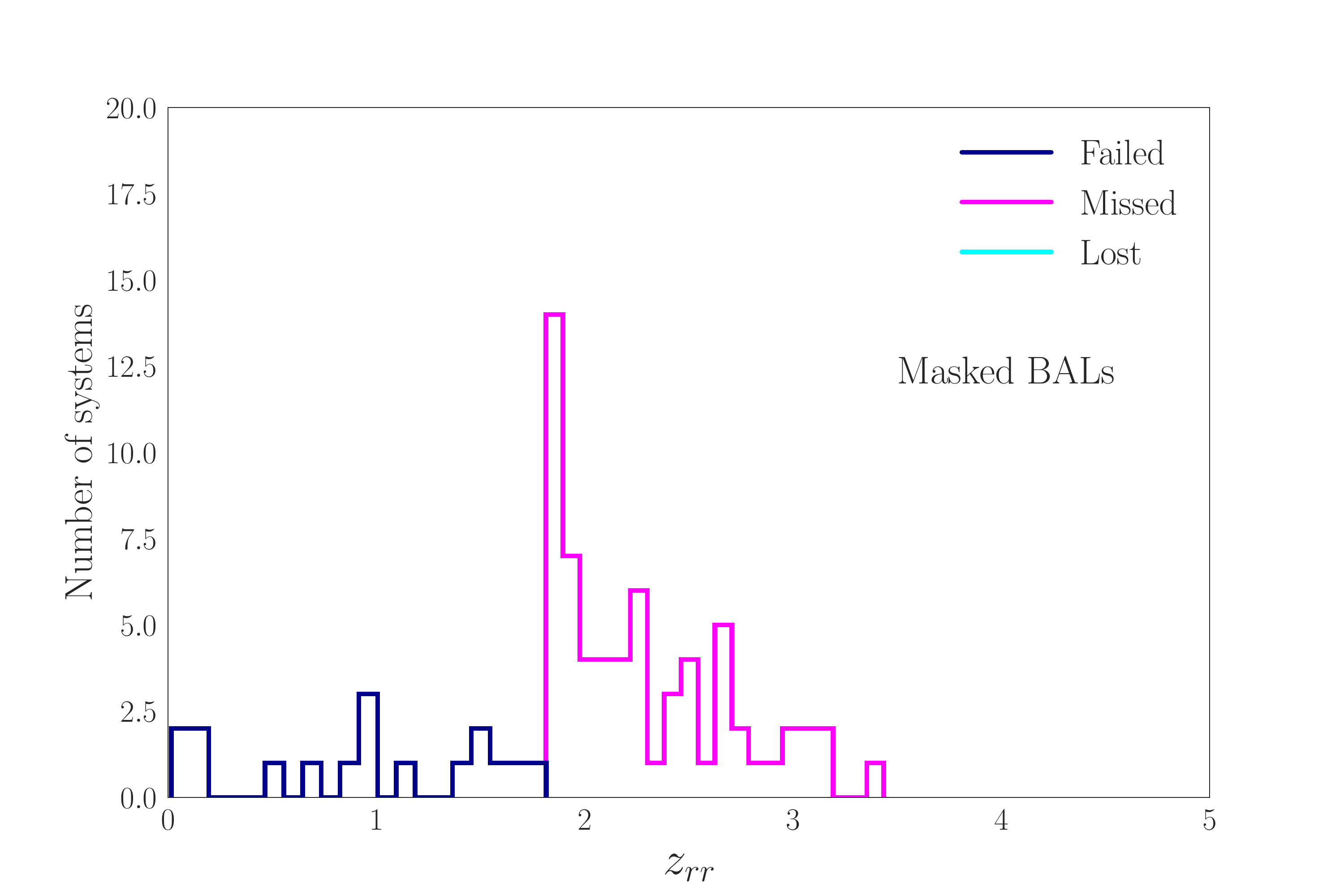}
\vspace{0.0cm}
\caption{\label{fig:ap_b} Distribution of failed fits, missed and lost opportunities as a function of the estimated redshift by \texttt{redrock} for different exposure times, which is a proxy for SNR. We compare synthetic spectra with unmasked BAL features and masked BAL features in the left and right columns, respectively. Conversely, the exposure times are presented in rows, from top to bottom: 1000 s, 2000 s, 3000 s, and 4000 s. Note that in the masked BAL case lost opportunities barely appear, which demonstrates that masking is improving \texttt{redrock}'s performance.}
\end{figure*}

We also re-calculate the fits, considering the different exposure times in the sample. Table~\ref{tab:xii} and Figure~\ref{fig:ap_b} show the results on this analysis while taking the difference between $z_{rr}$ and $z_{rr;noBAL}$.

\begin{table}
\centering
\caption{Success rates achieved by \texttt{redrock} for different exposure times.}
\label{tab:xii} 
\resizebox{0.48\textwidth}{!}{%
\begin{tabular}{|lcccc|}
\hline
\multicolumn{5}{|c|}{Exposure time $=$ 1000 s}  \\ \hline
\multicolumn{1}{|l|}{} & \multicolumn{1}{c|}{Good (\%)} & \multicolumn{1}{c|}{Failed (\%)} & \multicolumn{1}{c|}{Missed (\%)} & Lost (\%) \\ \hline
\multicolumn{1}{|l|}{$z_{rr;BAL} - z_{rr;noBAL}$} & \multicolumn{1}{c|}{96.24} & \multicolumn{1}{c|}{0.44}& \multicolumn{1}{c|}{3.30} & 0.02 \\ 
\multicolumn{1}{|l|}{$z_{rr;mas} - z_{rr;noBAL}$} & \multicolumn{1}{c|}{96.94} & \multicolumn{1}{c|}{0.08}& \multicolumn{1}{c|}{2.97} & 0.01 \\ \hline
\end{tabular}
}
\resizebox{0.48\textwidth}{!}{%
\begin{tabular}{|lcccc|}
\hline
\multicolumn{5}{|c|}{Exposure time $=$ 2000 s}  \\ \hline
\multicolumn{1}{|l|}{} & \multicolumn{1}{c|}{Good (\%)} & \multicolumn{1}{c|}{Failed (\%)} & \multicolumn{1}{c|}{Missed (\%)} & Lost (\%) \\ \hline
\multicolumn{1}{|l|}{$z_{rr;BAL} - z_{rr;noBAL}$} & \multicolumn{1}{c|}{98.06} & \multicolumn{1}{c|}{0.44}& \multicolumn{1}{c|}{1.49} & 0.01 \\ 
\multicolumn{1}{|l|}{$z_{rr;mas} - z_{rr;noBAL}$} & \multicolumn{1}{c|}{98.75} &
\multicolumn{1}{c|}{0.11}& \multicolumn{1}{c|}{1.13} & 0.01 \\ \hline
\end{tabular}
}
\resizebox{0.48\textwidth}{!}{%
\begin{tabular}{|lcccc|}
\hline
\multicolumn{5}{|c|}{Exposure time $=$ 3000 s}   \\ \hline
\multicolumn{1}{|l|}{} & \multicolumn{1}{c|}{Good (\%)} & \multicolumn{1}{c|}{Failed (\%)} & \multicolumn{1}{c|}{Missed (\%)} & Lost (\%) \\ \hline
\multicolumn{1}{|l|}{$z_{rr;BAL} - z_{rr;noBAL}$} & \multicolumn{1}{c|}{98.77} & \multicolumn{1}{c|}{0.44}& \multicolumn{1}{c|}{0.78} & 0.01 \\ 
\multicolumn{1}{|l|}{$z_{rr;mas} - z_{rr;noBAL}$} & \multicolumn{1}{c|}{99.26} &
\multicolumn{1}{c|}{0.12}& \multicolumn{1}{c|}{0.62} & 0.00 \\ \hline
\end{tabular}
}
\resizebox{0.48\textwidth}{!}{%
\begin{tabular}{|lcccc|}
\hline
\multicolumn{5}{|c|}{Exposure time $=$ 4000 s}\\ \hline
\multicolumn{1}{|l|}{} & \multicolumn{1}{c|}{Good (\%)} & \multicolumn{1}{c|}{Failed (\%)} & \multicolumn{1}{c|}{Missed (\%)} & Lost (\%) \\ \hline
\multicolumn{1}{|l|}{$z_{rr;BAL} - z_{rr;noBAL}$} & \multicolumn{1}{c|}{98.93} & \multicolumn{1}{c|}{0.43}& \multicolumn{1}{c|}{0.63} & 0.01 \\ 
\multicolumn{1}{|l|}{$z_{rr;mas} - z_{rr;noBAL}$} & \multicolumn{1}{c|}{99.41} &
\multicolumn{1}{c|}{0.12}& \multicolumn{1}{c|}{0.47} & 0.00 \\ \hline
\end{tabular}
}
\end{table}

\section{} \label{app:C}

In addition to the test presented in Appendix~\ref{app:B}, we assess the accuracy of our results by comparing the goodness of \texttt{redrock} fits with the same number of spectra in each exposure time (11546) in each realization in Table~\ref{tab:ap3}. \newline
\begin{table}
\centering
\caption{Success rates in fits achieved by \texttt{redrock}, when considering the possible exposure times.}
\label{tab:ap3} 
\resizebox{0.48\textwidth}{!}{%
\begin{tabular}{|lcccc|}
\hline
\multicolumn{5}{|c|}{Exposure time $=$ 1000 s}  \\ \hline
\multicolumn{1}{|l|}{} & \multicolumn{1}{c|}{Good (\%)} & \multicolumn{1}{c|}{Failed (\%)} & \multicolumn{1}{c|}{Missed (\%)} & Lost (\%) \\ \hline
\multicolumn{1}{|l|}{$z_{rr;BAL} - z_{rr;noBAL}$} & \multicolumn{1}{c|}{91.82} & \multicolumn{1}{c|}{0.99}& \multicolumn{1}{c|}{5.09} & 2.10 \\ 
\multicolumn{1}{|l|}{$z_{rr;mas} - z_{rr;noBAL}$} & \multicolumn{1}{c|}{95.15} & \multicolumn{1}{c|}{0.11}& \multicolumn{1}{c|}{4.14} & 0.60 \\ \hline
\end{tabular}
}
\resizebox{0.48\textwidth}{!}{%
\begin{tabular}{|lcccc|}
\hline
\multicolumn{5}{|c|}{Exposure time $=$ 2000 s}  \\ \hline
\multicolumn{1}{|l|}{} & \multicolumn{1}{c|}{Good (\%)} & \multicolumn{1}{c|}{Failed (\%)} & \multicolumn{1}{c|}{Missed (\%)} & Lost (\%) \\ \hline
\multicolumn{1}{|l|}{$z_{rr;BAL} - z_{rr;noBAL}$} & \multicolumn{1}{c|}{95.89} & \multicolumn{1}{c|}{1.11}& \multicolumn{1}{c|}{2.16} & 0.84 \\ 
\multicolumn{1}{|l|}{$z_{rr;mas} - z_{rr;noBAL}$} & \multicolumn{1}{c|}{98.69} & \multicolumn{1}{c|}{0.06}& \multicolumn{1}{c|}{1.19} & 0.06 \\ \hline
\end{tabular}
}
\resizebox{0.48\textwidth}{!}{%
\begin{tabular}{|lcccc|}
\hline
\multicolumn{5}{|c|}{Exposure time $=$ 3000 s}   \\ \hline
\multicolumn{1}{|l|}{} & \multicolumn{1}{c|}{Good (\%)} & \multicolumn{1}{c|}{Failed (\%)} & \multicolumn{1}{c|}{Missed (\%)} & Lost (\%) \\ \hline
\multicolumn{1}{|l|}{$z_{rr;BAL} - z_{rr;noBAL}$} & \multicolumn{1}{c|}{96.97} & \multicolumn{1}{c|}{1.24}& \multicolumn{1}{c|}{1.29} & 0.50 \\ 
\multicolumn{1}{|l|}{$z_{rr;mas} - z_{rr;noBAL}$} & \multicolumn{1}{c|}{99.29} & \multicolumn{1}{c|}{0.06}& \multicolumn{1}{c|}{0.64} & 0.01 \\ \hline
\end{tabular}
}
\resizebox{0.48\textwidth}{!}{%
\begin{tabular}{|lcccc|}
\hline
\multicolumn{5}{|c|}{Exposure time $=$ 4000 s}\\ \hline
\multicolumn{1}{|l|}{} & \multicolumn{1}{c|}{Good (\%)} & \multicolumn{1}{c|}{Failed (\%)} & \multicolumn{1}{c|}{Missed (\%)} & Lost (\%) \\ \hline
\multicolumn{1}{|l|}{$z_{rr;BAL} - z_{rr;noBAL}$} & \multicolumn{1}{c|}{97.34} & \multicolumn{1}{c|}{1.30}& \multicolumn{1}{c|}{0.94} & 0.42 \\ 
\multicolumn{1}{|l|}{$z_{rr;mas} - z_{rr;noBAL}$} & \multicolumn{1}{c|}{99.49} & \multicolumn{1}{c|}{0.07}& \multicolumn{1}{c|}{0.44} & 0.0 \\ \hline
\end{tabular}
}
\end{table}

Interestingly, Table~\ref{tab:ap3} reveals that the good fits are always above 90\% regardless of the exposure time considered. However, there is a higher success rate when BAL-QSOs are only 16\% of the total population of quasars, as explored in Table~\ref{tab:xii} compared with the results in this Appendix, in Table~\ref{tab:ap3}.

\bsp
\label{lastpage}
\end{document}